\documentclass[12pt,preprint]{aastex}

\usepackage{natbib,references}

\slugcomment{To be submitted to the Astrophysical Journal}
\shorttitle{Coronal Structure and Abundances in Young Fast Rotators}
\shortauthors{Garc\'{\i}a-Alvarez et al.}
\begin{document}
\title{Coronae of Young Fast Rotators}
\author{David Garc\'{\i}a-Alvarez\footnote{Current address: Imperial College London, Blackett Laboratory, Prince Consort Road, London SW7 2BZ}, Jeremy J. Drake,
  V.L. Kashyap. L. Lin, B. Ball} 
\affil{$^1$Harvard-Smithsonian Center for Aastrophysics, \\ 60 Garden Street, \\ Cambridge,
  MA 02138}

\begin{abstract}

AB~Dor, Speedy Mic and Rst~137B are in their early post-T Tauri
evolutionary phase ($\leq$ 100~Myr), at the age of fastest rotation in
the life of late-type stars.  They straddle the coronal
saturation-supersaturation boundary first defined by young stars in
open clusters.  High resolution {\it Chandra} X-ray spectra have
been analysed to study their coronal properties as a function of
coronal activity parameters {\it Rossby number}, $L_X/L_{bol}$ and a
coronal temperature index.  Plasma emission measure distributions as a
function of temperature show broad peaks at $T \sim 10^7$~K.
Differences between stars suggest that as supersaturation is reached
the DEM slope below the temperature of peak DEM becomes shallower,
while the DEM drop-off above this temperature becomes more pronounced.
A larger sample comprising our three targets and 22 active stars
studied in the recent literature reveals a general increase of plasma
at $T \ga 10^7$ toward the saturated-supersaturated boundary but a
decline beyond this among supersaturated stars.  All three of the
stars studied in detail here show lower coronal abundances of the low
FIP elements Mg, Si and Fe, relative to the high FIP elements S, O and
Ne, as compared to the solar mixture.  The coronal Fe abundances of
the stellar sample are inversely correlated with $L_X/L_{bol}$,
declining slowly with rising $L_X/L_{bol}$, but with a much more sharp
decline at $L_X/L_{bol}\ga 3\times 10^{-4}$.  For dwarfs the Fe
abundance is also well-correlated with Rossby number.  Coronal O
abundances appear lower than photospheric expectations by up to $\sim
0.2$~dex, but with no obvious trends with activity indices.  The
coronal O/Fe ratios for dwarfs show a clear increase with decreasing
Rossby number, apparently reaching saturation at [O/Fe]=0.5 at the
coronal supersaturation boundary.  Similar increases in O/Fe with
increasing coronal temperature and $L_X/L_{bol}$ are seen.  The range
in O/Fe variations attributable to outer atmosphere chemical
fractionation in our sample is about a factor of 10.

\end{abstract}


\section{Introduction}
In low-mass main sequence (MS) stars, internal structure is determined
primarily by stellar mass rather than age. In contrast, surface
activity as manifested in X-rays, at least for late-type dwarfs, seems
to scale directly with rotation and by consequence with age, but is
only weakly dependent on mass
\citep{Skumanich72,Hempelmann95,Stauffer97}.  
Observations suggest that, compared to the Sun, stars of higher
rotation rate show a more intense X-ray emission that reaches a
maximum of about $L_{\rm x}/ L_{\rm bol}=10^{-3}$ at rotation rates of
about $P_{rot}$=2-4.5d \citep{Stauffer97}. Beyond this rate lies the
"saturated'' regime where the X-ray luminosity is independent of
rotation.  This behaviour persists until rotation rates of about
$P_{rot}<$0.5d \citep{Prosser96,Randich98,Stepien01}, at which point
the X-ray luminosity is seen in open cluster stars to decrease
again. This regime is referred to as ``supersaturated''
\citep{Prosser96,Randich98}.

Although a number of different explanations have been invoked in order
to explain the saturation and supersaturation phenomenon
\citep[e.g.,][]{Jardine99,Jardine04,Ryan05}, there is as yet no widely
accepted theoretical descripton.  Existing studies pointing to
supersaturation are based on relatively crude diagnostics such as the
ratio of X-ray to bolometric luminosity, $L_x/L_{bol}$.  Issues such
as how coronal thermal structure or chemical composition anomalies
might change in the transition between the saturated and
supersaturated regimes have not yet been addressed.

{\it Chandra} and {\it XMM-Newton} high-resolution X-ray spectra allow
us to resolve spectral lines of a number of elements in the coronae of
active stars, and thus to probe these questions with greater acuity
than was possible with low resolution instruments.  
Despite the growing number of high-resolution X-ray
studies of late-type stars, there are as yet few studies on the
fastest-rotating 
so-called supersaturated stars.

In this paper, using {\it Chandra} High Energy Transmission Grating
spectrograph (HETGS) observations, we present a comparative analysis
of the coronal X-ray spectra of three late-type dwarfs of similar age
(probably 30-100~Myr; see \S\ref{s:stars}) rotating at, nominally, saturated and supersatured
rates. Our main aims are to
investigate differences in thermal structure and chemical composition
as a function of the ``Rossby number'' dynamo activity index
\citep{Durney78,Noyes84} in the supposedly saturated and
supersaturated regimes.

We first describe the three stars and briefly review earlier work
(\S\ref{s:stars}), then in \S\ref{s:obs} we report on the {\it
Chandra} observations and data reduction.  The methods used for a
differential emission measure analysis together with results obtained
are shown in \S\ref{s:anal}. In \S\ref{s:results} and \S\ref{s:lite} 
we discuss our results on the coronal abundances and temperature structure 
and report our conclusions in \S\ref{s:concl}. Note that the results for AB~Dor
we present here were reported by \citet{Garcia-Alvarez05} to which we
refer the reader for further details.

\section{Program Stars}
\label{s:stars}

Our sample stars have been chosen to 
 represent late-type dwarfs at an early evolutionary phase
($\leq$ 100~Myr) during which they attain their fastest rotation rates.

\subsection{AB~Dor}
\label{s:abdor}

AB~Doradus (HD 36705) is a young, relatively bright (V=6.9) and
rapidly rotating late-type star with spectral type K0-2\,V. It is an
example of the very active cool stars that are just evolving onto the
main sequence. In recent years there have been a controversy over the 
age of AB~Dor with two rather different estimates for its age: 
a first one of the order of 50 Myr (e.g.; \citet{Zuckerman04} $\sim$50~Myr; 
\citet{Lopez-Santiago06} $\sim$50~Myr; \citet{Janson07} 50-100~Myr) 
and a second one of about 100-125Myr (e.g.; \citet{Luhman05} 75-125~Myr; 
\citet{Ortega07} 100-140~Myr). The main physical parameters of AB~Dor are
shown in Table~\ref{t:par}. AB~Dor has been extensively observed in
the ultraviolet \citep{Rucinski95,Moos00,Vilhu01} and X-rays
\citep{Pakull81,Vilhu87,Collier-Cameron88,Mewe96,Kuerster97,Maggio00,Guedel01}.

The photospheric abundances of AB~Dor and its hosting cluster, the
Pleiades, are relatively well studied
\citep[e.g.;][]{Vilhu87,King00,Boesgaard89b,Cayrel85,Schuler04}. All
these studies reported a solar-like photospheric mixture.

A low coronal iron abundance, [Fe/H]=-0.95, found by \citet{Mewe96}
using simultaneous observations with EUVE and ASCA, was confirmed by
\citet{Guedel01} using {\it{XMM-Newton}} data. \citet{Sanz-Forcada03}
reported coronal abundances for AB~Dor based on XMM-Newton and {\it
Chandra} spectra showing a similar Fe
depletion. \citet{Garcia-Alvarez05} found similar results for AB~Dor
but no evidence for enhancements of very low FIP ($< 7$~eV) elements,
such as Na, Al and Ca, suggested by previous works.

\subsection{Rst~137B}
\label{s:rst137B}

Rst~137B is a common-proper-motion companion of AB~Dor, at a
separation of $~10''$, with spectral type M3-M5.  Rst~137B was first
discovered in X-rays from an {\it Einstein} High Resolution Imager
observation \citep{VilhuLinsky87}.  It is a strong X-ray and UV
source, with the observed emission at both wavelengths lying close to
the saturation levels defined by young and/or rapidly rotating stars
\citep{VilhuLinsky87,Vilhu89}.  \citet{Martin95} confirmed, based on the
existing data (space velocities, rotation, emission line strengths,
lithium) the evolutionary status intermediate between T Tauri stars
and Pleiades stars for Rst~137B.  \citet{Collier-Cameron97} concluded
that the relative magnitudes and V(RI)$_c$ colours of AB Dor and
Rst~137B are consistent with both stars having a common evolutionary
age that is no greater than that of the Pleiades. This age makes
Rst~137B to be an order of magnitude younger than the majority of
flare stars in the solar neighborhood.  Further evidence for its youth
are provided by its intense \ion{Ca}{2} emission \citep{Innis88}, high
X-ray luminosity of log~L$_x$/L$_{bol}\sim$-3 \citep{VilhuLinsky87},
and high projected angular velocity \citep[v sin i= 50
$km\,s^{-1}$][]{Vilhu91}, which is comparable to that of the most
rapidly rotating M dwarfs in the Pleaides cluster \citep{Stauffer87}.
\citet{Lim93} estimated a rotational 
period of $P_{\mathsf{rot}}<9hr$ based on projected rotation velocity
and spectral type.  Important  physical parameters of Rst~137B are listed
in Table~\ref{t:par}.

\subsection{Speedy Mic}
\label{s:speedymic}

Attention was drawn to BO Mic by \citet{Bromage92}, who reported what
turned out to be the largest stellar flare observed during the EUV
all-sky-survey of ROSAT. This single young K-dwarf is one of the most
active solar neighbourhood stars \citep{Singh99}, with an X-ray to
bolometric 
luminosity ratio of $\log(L_X/L_{bol})$ =-3.07.  It can reach a value of
$\log(L_X/L_{bol})\sim -2$ during flares \citep{Makarov03}. BO Mic is
usually nicknamed ``Speedy Mic'' 
due to it fast rotation ($P_{\mathsf{rot}}=0.380\pm0.004$ days
\citep{Cutispoto97}, v sin i= 132 $km\,s^{-1}$ \citep{Barnes05}). The
main physical parameters of Speedy~Mic are listed in
Table~\ref{t:par}. \citet{Montes01} have shown that Speedy Mic, like
AB Dor, is indeed a member of the Local Association. \citet{Barnes01}
confirmed the presence of H$\alpha$ transients, first reported by
\citep{Jeffries93}. These transients are thought to be the result of
clouds of cool material, analogous to solar prominences passing 
in front of the stellar disc.
\citet{Singh99} found significantly subsolar coronal Fe abundances,
based on {\it ASCA} observations, in accordance with observations of
other active stars. \citet{Nordstrom04} derived a solar-like
metallicity ([Fe/H]=0.03) for Speedy Mic from photometric indices
(since the same study found AB~Dor to be depleted in Fe by a factor of
three compared to solar and in contradiction to the solar composition
found from high resolution spectroscopy we place only limited weight
on this result).

\section{Observations}
\label{s:obs}

The {\it{Chandra}} HETGS observation of AB~Dor, Speedy~Mic and
Rst~137B were carried out using the Advanced CCD Imaging Spectrometer
(ACIS-S). All the observations employed the detector in its standard
instrument configuration. Rst~137B was observed simultaneously and
serendipitously during the AB~Dor observations.  Two observations one
day apart were obtained for Speedy~Mic. These were summed resulting in
an exposure time of 70ks.  The observations of the three targets are
summarized in Table~\ref{t:par}.

\begin{figure}
\plotone{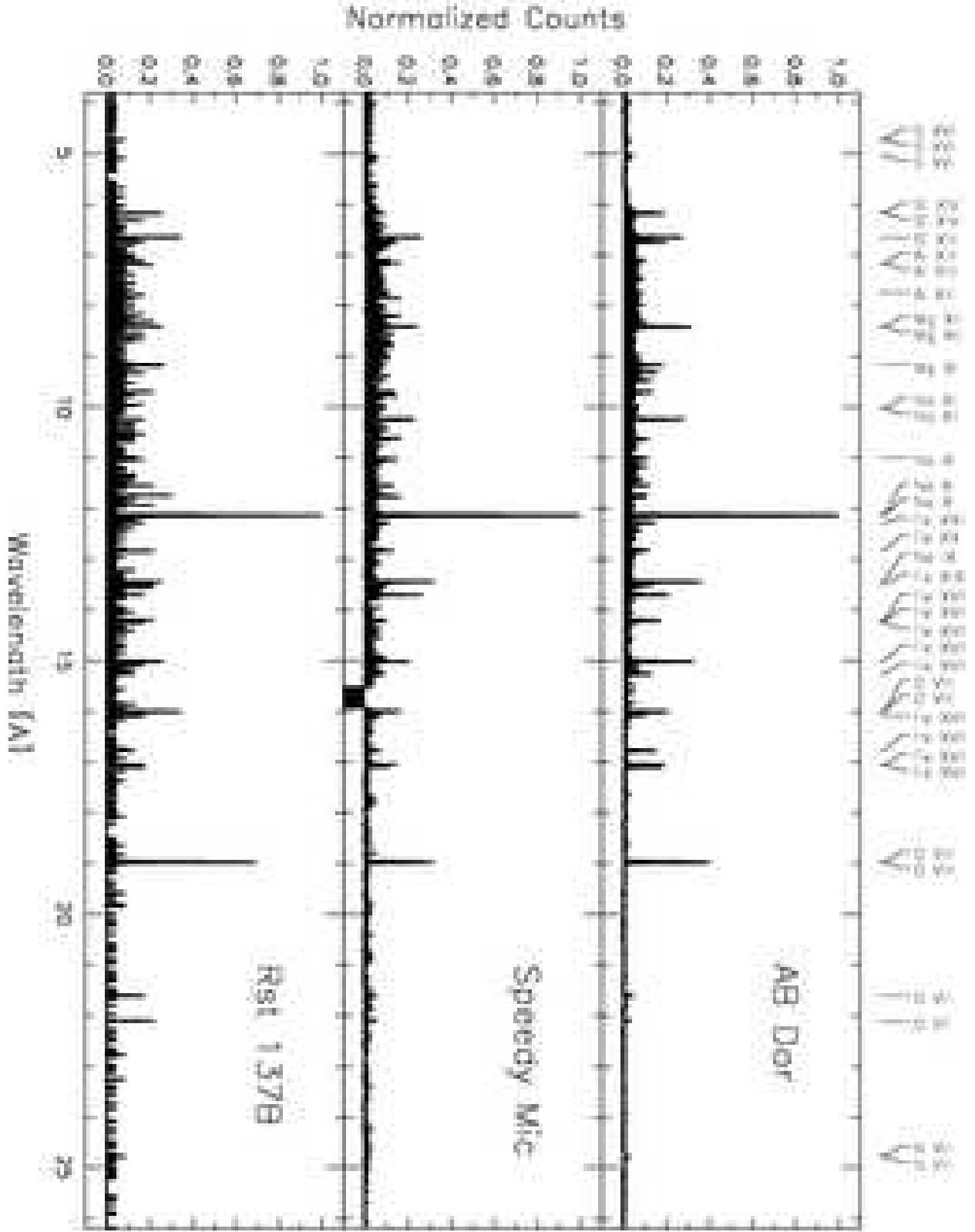}
\caption{{\it Chandra} X-ray spectra of AB~Dor, Speedy Mic and
  Rst~137B. The strongest lines over the observed wavelength range are
  identified.} 
\label{f:sp}
\end{figure}

Fig.~\ref{f:sp} shows the {\it Chandra} X-ray spectra of AB~Dor,
Speedy~Mic and Rst~137B in the wavelength range 4-26~\AA, which
contains the prominent lines of N, O, Ne, Na, Mg, Al, Si, S and Fe. 
The strongest coronal lines are identified. These spectra show a
remarkable similarity of lines, both in terms of which lines are
prominent, from H- and He-like ions and the broad range of charge
states of Fe, and in their relative intensities. \ion{Ne}{10} is the
strongest line observed in the spectrum of the three targets.  The
spectrum of Rst~137B, although similar to the other two stars, shows
comparatively stronger O and Si lines. We also observe that in
Speedy~Mic and Rst~137B the Fe lines are comparatively weaker than
their counterparts in AB~Dor.

\section{Analysis}
\label{s:anal}
Pipeline-processed (CXC software version 6.3.1) photon event lists
were reduced using the CIAO software package version 3.2, and were
analyzed using the IDL\footnote{Interactive Data Language, Research
Systems Inc.}-based PINTofALE\footnote{Available from http://hea-www.harvard.edu/PINTofALE} 
software suite \citep{Kashyap00}. The analysis we have performed consisted
of line identification and fitting, reconstruction of the plasma
emission measure distribution including allowance for blending of the
diagnostic lines used, and finally, determination of the element
abundances.
\subsection{Photometry}

\begin{figure}
\plotone{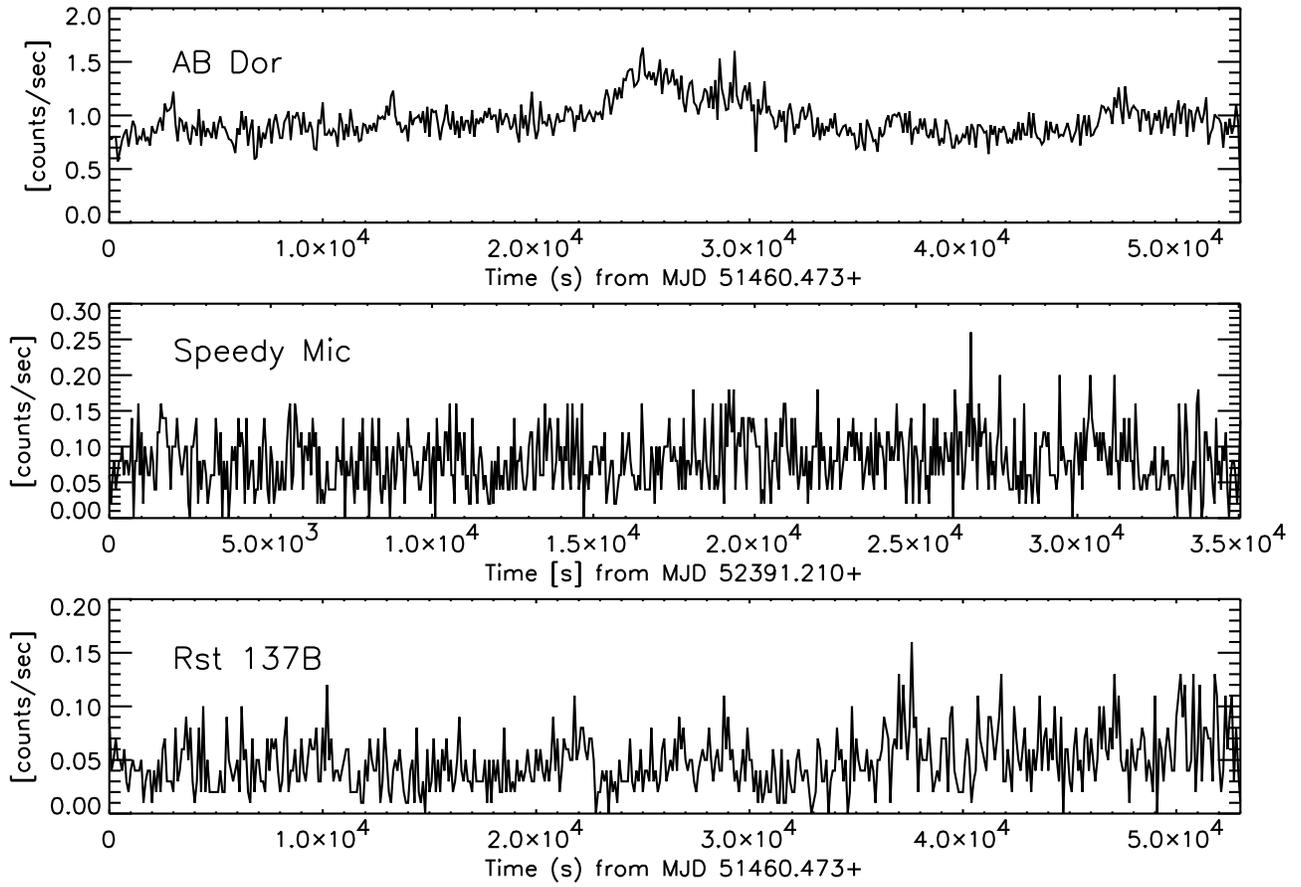}
\caption{{\it Chandra} X-ray light curves of AB~Dor, Speedy Mic and
  Rst~137B binned at 100s intervals. The three objects were relatively
  quiescent, showing no large flare events, excepting the moderate
  event on AB~Dor midway through the observation.} 
\label{f:lc}
\end{figure}

Before commencing spectral analysis, we first checked for flare
activity that could affect not only the shape of the differential
emission measures (DEMs) but might also be accompanied by detectable
changes in the chemical composition of the plasma that dominate the 
disk-averaged spectra \citep[e.g.;][]{Favata00,Maggio00,Guedel01}.
Light curves for AB~Dor, Speedy~Mic and Rst~137B observations were
derived from the dispersed photons only and excluded the 0th order,
which could be affected by pileup. Events were then binned at 100s
intervals.  The resulting light curves are illustrated in
Fig.~\ref{f:lc}. Note that all the light curves are relatively flat
and devoid of significant flare activity, excepting the moderate event
on AB~Dor midway through the observation. We conclude that AB~Dor,
Speedy~Mic and Rst~137B observations are representative of the stars
during times of relative quiescence, and therefore treat the observations in
their entirety for the remainder of the analysis.

\subsection{Spectroscopy}

Spectral line fluxes for AB~Dor, Speedy~Mic and Rst~137B were measured
by fitting modified Lorentzian or Moffat (``beta profile'') functions
which represents the {\it{Chandra}} transmission grating instrumental
profile to within photon counting statistics for lines of the order of
a few 1000 counts or less \citep{Drake04b}. Table~\ref{t:flx} shows
the measured fluxes of the emission lines identified and used in this
analysis.

\subsection{Differential Emission Measure and Coronal Abundances}
\label{s:dems}

Our method for obtaining the {\em differential emission measure} (DEM)
has been described in detail in earlier works \citep{Garcia-Alvarez05}
and is described here only in brief.  A given line flux depends on
plasma temperature and on the abundance of the element in question.  The
ratio of two emission lines from ions of the {\em same} element is
independent of the abundance of the chosen element.  We have therefore
devised a method that uses line {\em ratios} instead of line fluxes
directly.

We use a Markov-Chain Monte-Carlo analysis using a Metropolis
algorithm (MCMC[M]) on the set of supplied line flux ratios
\citep{Kashyap98} to estimate the DEM. Our basic set of diagnostics
comprises the H-like/He-like resonance line flux ratios for the
elements O, Ne, Mg, and Si, line ratios involving Fe~XVII, Fe~XVIII
and Fe~XXI resonance lines, and measurements of the continuum flux at
points in the spectrum that are essentially free of lines. These
continuum values serve both to constrain the emission at high
temperatures and to determine the absolute normalization of the
DEM. The set of lines include the brightest lines in stellar coronal
spectra and are easily measured in essentially all well-exposed {\em
Chandra} grating observations of stellar coronae, such that
star-to-star variations in diagnostic lines used can be avoided. The
emission measure distribution and abundance analysis employed the
CHIANTI database version 4.2 \citep{Dere01} and the ionization balance
of \citet{Mazzota98}, as implemented in the PINTofALE software package
\citep{Kashyap00}. Hidden blends were corrected for based on the
prediction of blending lines in the CHIANTI database using the methods
described in \citet[][]{Garcia-Alvarez05}.

The MCMC[M] method yields the emission measure distribution over a
pre-selected temperature grid, where the DEM is defined for each T
bin.  In our case, a set of temperatures $T_n$, with
$\Delta\,\log\,T[K]$=0.1, define the
DEM(T$_n$)=${n_e^2}(T_n)\frac{dV(T_n)}{dlogT}$. The derived DEM(T$_n$)
is reliable over a certain temperature range only if we have 
lines with contribution function G($T_{max}$)$\sim$ G($T_{n}$). Based
on the lines we use in our analysis we are able to obtain a
well-constrained DEM(T$_n$) between $\log\,T[K]$=6.2 (the coolest peak
formation temperature given by the resonance line \ion{O}{7} is
$\log\,T[K]$=6.3) and $\log\,T[K]$=7.5 (the hottest peak formation
temperature given by the resonance line \ion{S}{16} is
$\log\,T[K]$=7.4); larger uncertainties are obtained outside that
range. Our final DEMs are reported as volume emission measures at the
star. We have adopted distances $D$=14.9\,pc, $D$=44.4\,pc and
$D$=14.9\,pc for 
AB~Dor, Speedy~Mic and Rst~137B respectively \citep{Perryman97}.

\subsection{Abundances}
\label{s:abund}

Once the DEM has been established, we can evaluate the abundances of
any elements for which we have lines with measured fluxes.  We derived
values for the coronal abundances of the elements O, Ne, Mg, Si, S,
and Fe.  Na was also studied although only upper limits were obtained 
owing to the weakness of the line features within our spectral
bandpass; 
S was also not detected in Speedy~Mic and
RST~137B.  In cases where there was more than one line for a given 
element, the weighted average of the abundance
determinations from each of the individual lines was adopted.

According to previous studies in the literature, the photospheric
abundances for AB~Dor 
and Speedy~Mic are similar to the solar ones (see \S\ref{s:abdor} and
\S\ref{s:speedymic}). There are no photospheric abundance measurements
available for Rst~137B. However, its physical association with AB~Dor
(see \S\ref{s:rst137B}) presents a very plausible case for its
abundances being solar-like. We therefore adopt the mixture of
\citet{Asplund05} for comparing all coronal abundance results. The
\citet{Asplund05} solar chemical composition shows a downward revision
by 25-35\%\ of the abundances of light elements such as C, N, O and Ne
compared with values from earlies studies \citep[e.g.][]{Grevesse98}.
The use of these new values would increase the derived stellar coronal
abundances of light elements relative to those obtained by using the
solar chemical composition from earlies studies. In other words, the
observed trends in any coronal abundance FIP and inverse-FIP effects
\citep{Brinkman01} will
be slightly shallower and slightly steeper respectively if one uses
the new solar chemical composition reported by \citet{Asplund05}. One
exception to the \citet{Asplund05} mixture might be the element
Ne. \citet{Drake05b} have recently found Ne/O to be enhanced by an
average factor of 2.7 compared to the \citet{Asplund05} ratio in the
coronae of approximately 20 nearby and mostly active 
stars. Based on these results, we
have assumed a Ne abundance revised upward by 0.43\,dex with respect
to \citet{Asplund05}. We discuss this in more detail in \S\ref{s:abundances}.

We also used the temperature-insensitive abundance ratio diagnostics
of \cite{Drake07} as an additional check on our values obtained using
the DEM.  These are ratios formed by combining two sets of lines of
two different elements, constructed such that the combined emissivity
curves of each set have essentially the same temperature
dependence. The resulting ratio of measured line fluxes then yields
directly the ratio of the abundances of the relevant elements,
independent from the atmospheric temperature structure. Results are
shown in Table~\ref{t:abund2}. 

\section{Results}
\label{s:results}

\begin{figure}
\plotone{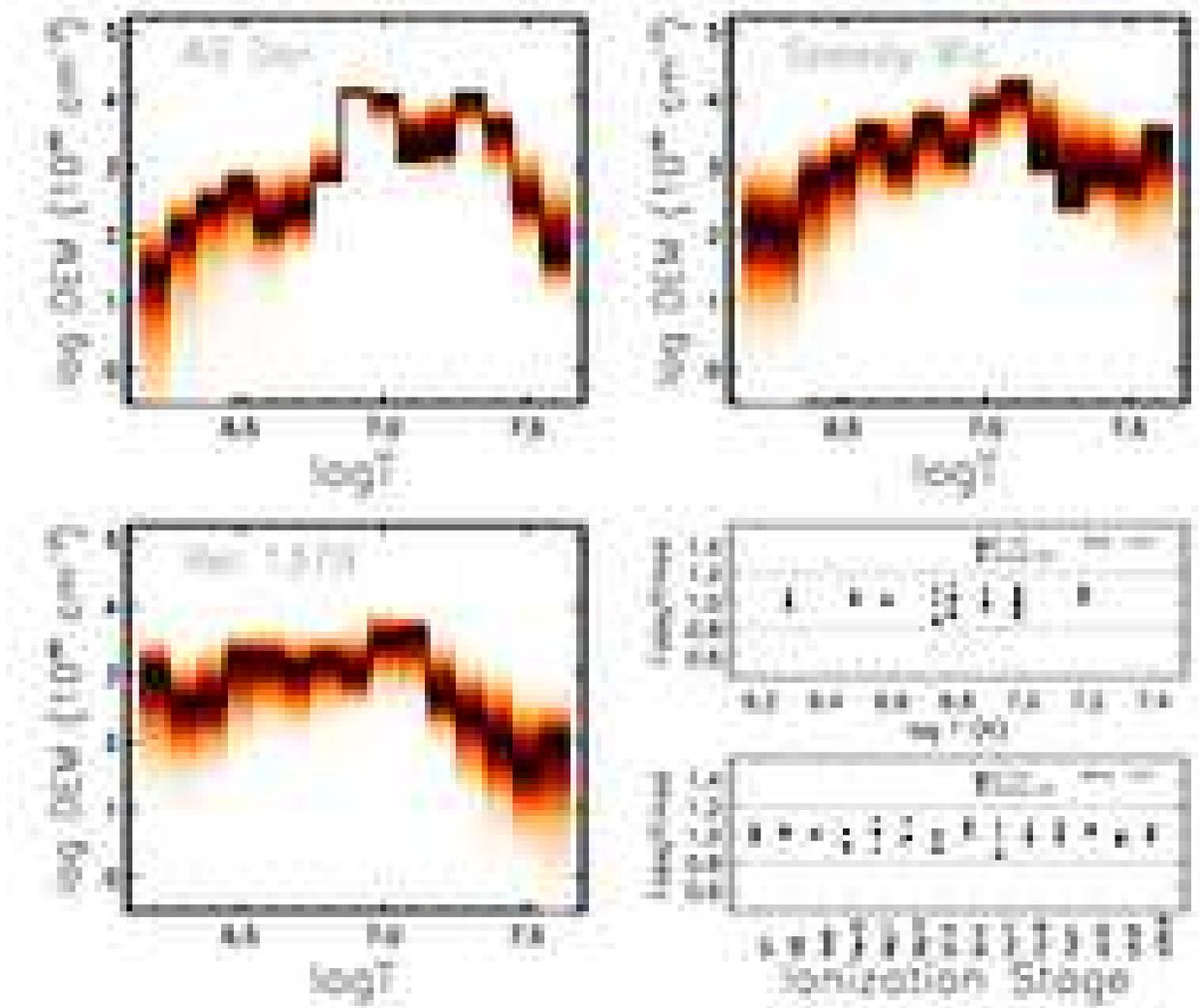}
\caption{Top and bottom left: DEMs obtained for AB~Dor, Speedy Mic and
  Rst~137B by running a 
MCMC[M] reconstruction code on a set of lines of
H-like, He-like and highly ionized Fe line fluxes (O, Ne, Mg, Si,
\ion{Fe}{17}, \ion{Fe}{18} and \ion{Fe}{21}). The thick solid line
represents the best-fit DEM, while the shaded regions correspond to
the 1-$\sigma$ deviations present in each temperature bin. Bottom
right: Comparison of observed and modelled line fluxes vs ionic 
species (bottom) and vs $T_{max}$ (top) for the three objects. The
dashed lines represent 1-$\sigma$ deviation.\label{fig:dem}}
\label{f:dems}
\end{figure}

In order to verify the propriety of our DEM and abundance techniques
we have compared observed and modeled line fluxes vs ionic species and
$T_{max}$ (the temperature of maximum emissivity) in Fig~\ref{f:dems}
(bottom right panel). All the predicted line fluxes based on our final
models are within 10~\%\ or so of the observed values.

\subsection{Temperature Structure}
\label{s:structure}

Fig.~\ref{f:dems} illustrates the reconstructed DEMs for our three stars
AB~Dor, Speedy~Mic and Rst~137B.  Although the DEMs for the three
targets peak around $\log\,T$[K]$\sim $7.0-7.1, the overall shapes are
slightly different.  The DEM for AB Dor shows a primary peak at
$\log\,T\sim 7.0$ that extents to $\log\,T\sim 7.4$. The DEM for
Speedy~Mic has a less pronounced peak around $\log\,T\sim 7.0$--7.1 and
shows a sharper decrease for $\log\,T> 7.1$. Finally, the DEM
for Rst~137B is relatively flat, peaking at $\log\,T\sim 7.0$--7.1 and
showing a smooth steep decrease for higher temperatures.

\subsubsection{Comparison with Previous Work}

As noted by \citet[][]{Garcia-Alvarez05}, the DEM for AB Dor is
comparable with that reported recently by \citet{Sanz-Forcada03} based
on XMM-{\it{Newton}} and {\it{Chandra}} observations. These authors
obtained a slightly smoother increase in emission measure with rising
temperature for $\log\,T< 6.6$, but a similar peak at $\log\,T\sim 7.0$
and sharp decrease for the higher temperatures, $\log\,T> 7.4$.
Similar DEM shapes have been derived based on EUVE and ASCA
observations of AB~Dor obtained in 1993 November 
by \citet{Rucinski95} and \citet{Mewe96}; these authors
also note that the results are quite sensitive to the coronal
abundances adopted.    

\citet{Singh99} estimated the DEM for Speedy Mic using a 6th
order Chebyshev polynomial method applied to an {\it ASCA}
observation.  While {\it ASCA} spectra alone do not possess much
temperature resolution, the single broad DEM peaking at $\log T\sim
7.1$ and also showing some evidence for plasma up to temperatures
$\log T > 7.7$, is qualitatively similar to our result.

To our knowledge, there exist no other DEM analyses
of Rst~137B by other authors with which to compare our temperature
structure results.  In general, the very limited information from
earlier observations indicates that the DEMs themselves are not
subject to large secular changes for these active stars.

There are no striking features to the DEMs of any of these stars that
distinguish them from those of similar fast rotators.  Other examples,
such PZ~Tel (rotation period $P=0.94$d) and YY
Gem ($P=0.81$d), show DEMs that peak at $\log T\sim 7.1$ with and
extension indicating significant plasma at temperatures a factor of 2
or so higher \citep{Stelzer02,Argiroffi04}.  These characteristics are
slightly more similar to those of AB~Dor than Speedy~Mic and Rst~137B.

\subsection{Coronal Abundances}

\begin{figure}[t]
\plotone{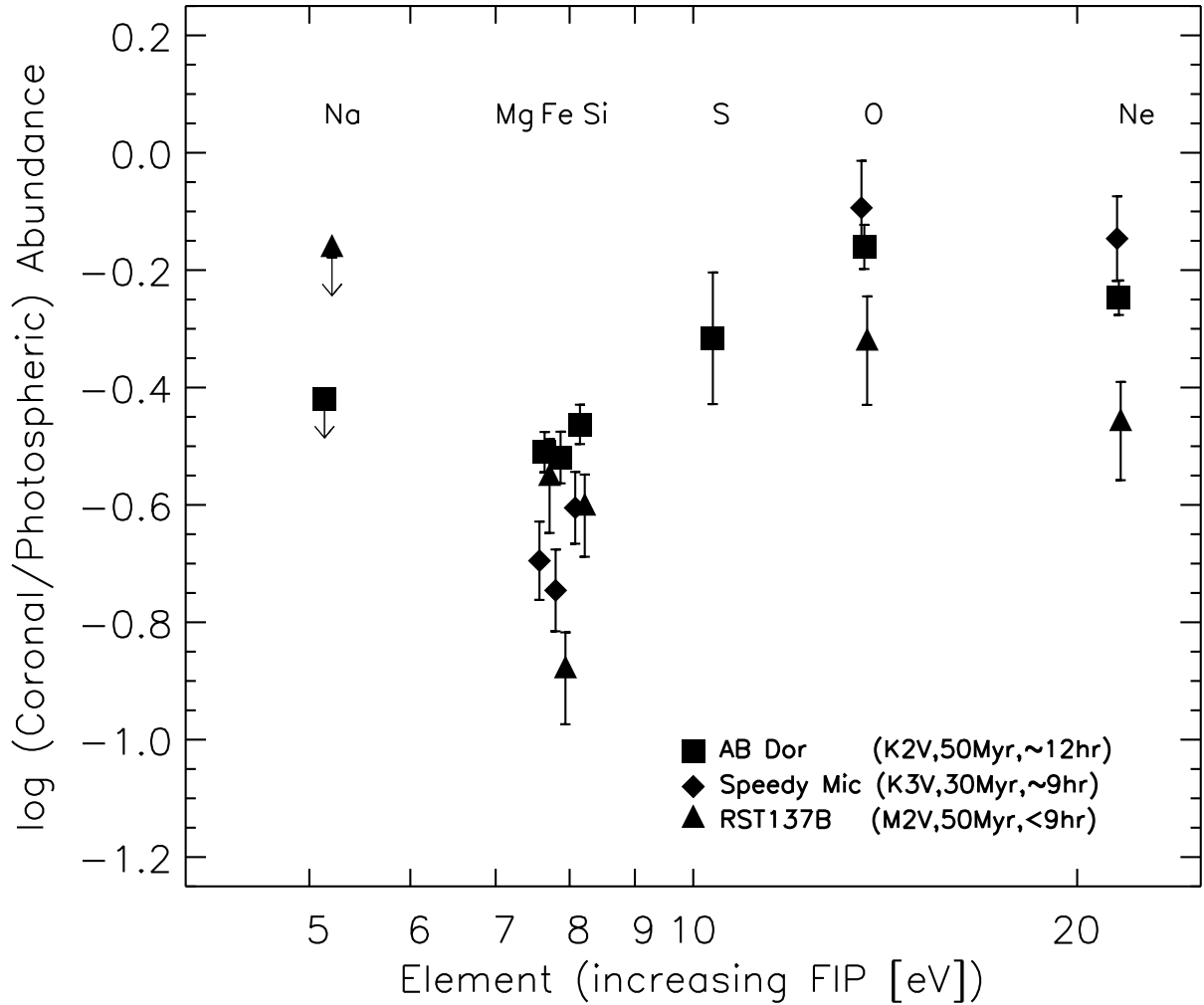}
\caption{{Comparison of coronal abundances vs FIP for AB~Dor,
    Speedy~Mic and Rst~137B. The abundances were obtained from the
    abundance-independent DEM reconstruction and are relative to the
    solar photospheric mixture of \citet{Asplund05} with the Ne from
    \citet{Drake05b}. True uncertainties in the coronal abundances are likely to be
    0.1\,dex.}. Details on methods used for the formal error estimates are given by \citet{Garcia-Alvarez05}.} 
\vspace{.6cm}
\label{f:abund}
\end{figure}

Our results for the element abundances of AB~Dor, Speedy~Mic and
Rst~137B are summarized in Table~\ref{t:abund1}. We only list
statistical uncertainties here.  Details on methods used for the
formal error estimates are given by \citet{Garcia-Alvarez05}. The
abundance ratios derived from temperature-insensitive line ratios (see
\S\ref{s:abund}) are listed in Table~\ref{t:abund2}.  Within expected
uncertainty ranges these values are
in agreement with, and provide verification for, the ones derived from
the DEMs (Table~\ref{t:abund1}), with the exception of the ratios
involving Fe. As noted by \citet{Drake07}, however, these ratios are
likely to exhibit larger systematic errors owing to a less optimum
coincidence of the different line contribution functions as a function
of temperature. In Fig.~\ref{f:abund} we have plotted the derived
abundances, relative to the adopted stellar photospheric values
(essentially the mixture of \citet{Asplund05}, with the exception of
Ne for which we adopt the \citet{Drake05b} value, as discussed
earlier), in order of element FIP.

All three stars show some evidence for what has become know as an
``inverse-FIP effect'', with depletion of the low FIP elements
($<$10\,eV) relative to photospheric values
\citep[e.g.][]{Brinkman01}.  We also note a slightly lower Ne
abundance with respect to the Ne/O ratio of \citet{Drake05b} for all
three stars; this is most pronounced for Rst~137B and AB~Dor.  It can
be seen from comparison of the Ne/O ratios obtained from DEM modelling
and temperature-insensitive line diagnostics (Table~\ref{t:abund2})
that the former method tends to arrive at a slightly lower Ne
abundance than the latter.  Uncertainties in both methods are of order
0.1~dex \citep[see also][]{Drake05}, and the current discrepency is
not beyond the bounds of the systematic errors of the two approaches.
A comparison between the results from these methods in general is
given in \citet{Garcia-Alvarez05}.

The coronal abundances derived for AB~Dor are in good agrement with
those reported in the literature \citep[see][ and references
therein]{Garcia-Alvarez05}.  The coronal abundances derived for
Speedy~Mic are in reasonable agrement with the only ones reported in
the literature for this object \citep{Singh99}.  The latter values
were obtained using low spectral resolution {\it ASCA}
observations. To our knowledge, there are no other coronal
abundance studies of Rst~137B for comparison.

Despite Mg, Fe and Si having very similar FIP, Speedy Mic and Rst~137B
show much lower Fe abundances than Si and Mg compared with
AB~Dor. This effect has also been seen in other coronal abundance
studies, as summarised by \citet{Drake03a} who suggested gravitational
settling of the heavier Fe ions as a possible explanation.  We discuss
this further below.

\section{Coronal Temperature Structure and Chemical Composition 
in the Saturation-Supersaturation Regime} 
\label{s:lite}

\subsection{Temperature Structure}

It has been especially emphasised in the context of the solar corona
that derived DEMs can be quite sensitive to the lines used for their
derivation and to the uncertainties in the atomic data used.  The best
way to investigate systematically any trends in temperature structure
with activity is therefore to use DEMs derived from observations made
with the same instrumentation and line set.  Our sample of three stars
severely limits any such discussion, though two possible trends are
apparent.  

The slopes in the derived DEMs on the low temperature side of their
maxima follow very approximately the relation DEM$\propto T^{3}$,
DEM$\propto T^{2}$ and DEM$\propto T$ for AB Dor, Speedy~Mic and
Rst~137B respectively.  This trend is in the direction of decreasing
rotation period, and Rossby number (see below).  Furthermore, it is
apparent that the the amount of plasma present at $\log T > 7$ falls
off in the same order.  Combined, the impression is that the ratio of
material at hot ($\log T> 7$) and cooler ($\log T < 7$) temperatures
tends to decrease with increasing saturation.

In order to compare these sparse, but systematically derived, results
with those of other stars, we have computed the temperature-sensitive 
index $\Phi_{6.9}$, which
we define as the ratio of the DEM in
the ``high'' temperature range ($\log\,T$=6.9-7.6) to the total DEM
($\log\,T$=6.2-7.6), for a larger sample of rapid rotators culled from
analyses in the literature.
While a fairly coarse representation of DEM, this temperature index
should be quite insensitive to the details of the different methods
and data used to derive it.  The sample of stars investigated is
listed in Table~\ref{t:sample}, together with their relevant properties.

Since we are studing stars of different spectral type, it is also
instructive to use the {\em Rossby number}---the ratio of the
rotational period to the convective turnover time $P/\tau_c$---instead
of the rotational period in order to compare the behaviour of their
coronal structure and chemical composition.  \citet{Durney78}
showed that the Rossby number is closely related to the dynamo number
for a fluid and is expected to scale directly with the generation of
magnetic flux.  \citet{Noyes84} first demonstrated the tight
correlation between the Ca~II H and K chromospheric activity
indicators and stellar Rossby number, and the correlation also holds
through to coronal activity indicators such as X-ray luminosity
\citep[see, e.g.][]{Randich98}.
The ratio $P/\tau_c$ is a function of stellar mass, and saturation will
start to be seen at progressively longer rotation periods for stars with
lower masses.  According to \citet{Randich98}, supersaturated stars
from open cluster data have Rossby numbers log$N_R$ $<-1.7$
\citep{Randich98}.  Therefore, Speedy~Mic and Rst~137B are in
principle included in this range, while AB Dor sits near the
saturated-supersaturated boundary region (see Table~\ref{t:par}).

\begin{figure}[t]
\plotone{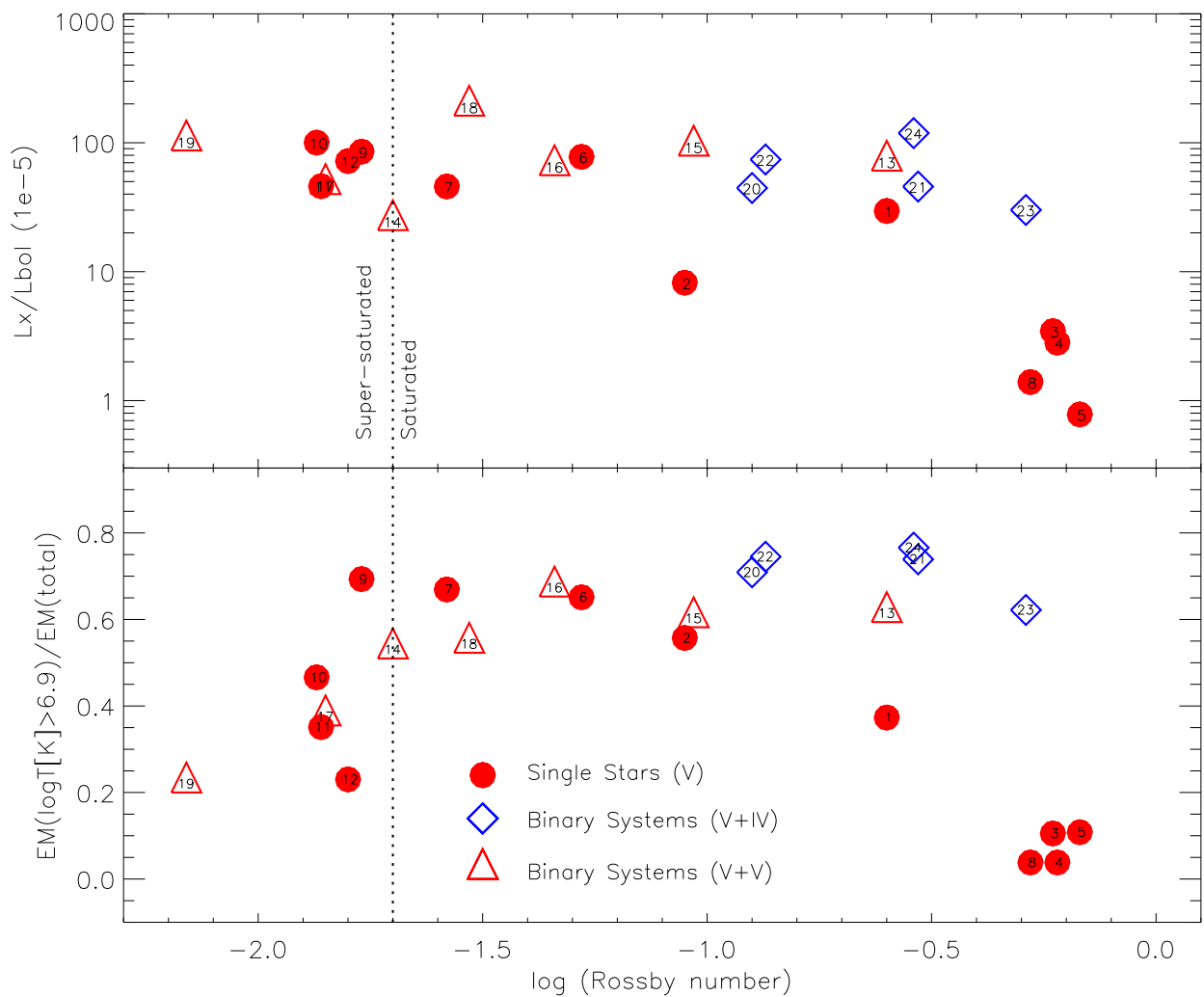}
\vspace{.6cm}
\caption{Top panel: $L_x/L_{bol}$  vs the Rossby number for the sample of stars in Table~\ref{t:sample}. This sample includes single dwarfs stars (circles) and binary systems (diamonds and triangles). The demarcation between the saturated and supersaturated regions, based on definition of \citet{Randich98}, is indicated by a dashed vertical line. Bottom panel: Same as for top panel but for $\Phi_{6.9}$ vs the Rossby number.} 
\label{f:ross_hilo}
\end{figure}

In Fig~\ref{f:ross_hilo} we show both $L_x/L_{bol}$ (upper panel) and
$\Phi_{6.9}$ (lower panel)
vs Rossby number for the sample of stars in Table~\ref{t:par}.
The $L_x/L_{bol}$ values for this sample do not exhibit any evidence
for the supersaturation effect, although a very clear saturation
plateau at the canonical value $L_x/L_{bol}\sim 10^{-3}$ is readily
apparent.  This lack of supersaturation signature in field stars was
pointed out by \citet{Randich98}, and also characterises the sample of
young cluster and field stars studied recently by
\citet{Jeffries05}.  Thus, while Speedy~Mic and Rst~137B are
within the regime of supersaturation, their $L_x/L_{bol}$ indices are
representative of the canonical saturated regime.

Despite the lack of supersaturation signature in $L_x/L_{bol}$, the
lower panel of Fig~\ref{f:ross_hilo} indicates different behaviour in
thermal structure.  As the Rossby number decreases and approaches the
saturated-supersaturated boundary region, the emission measure at the
``high'' temperature range increases relative to the total emission
measure.  However, once the supersaturated region is reached this
trend inverts, showing a decline in the emission measure of plasma at
high temperatures.  Thus, the supersaturation boundary that is
apparent through a decline in $L_x/L_{bol}$ past a critical Rossby
number of $\log N_R \sim -1.7$ in some open clusters
\citep{Prosser96,Randich98}, but appears not to be present in field
stars, does seem to be marked in the latter sample by this transition
to a somewhat cooler dominant corona.

\citet{Randich98} suggested, as a possible explanation for
supersaturation in L$_x$, that very rapid rotation could lead to a
substantially higher coronal temperature and to the shift of the DEM
out of the ROSAT passband.  However, our DEM results for AB~Dor,
Speedy~Mic and Rst~137B, in addition to the $\Phi_{6.9}$ indices for the sample
of stars in Table~\ref{t:sample}, show little evidence for substantial
amounts of plasma at temperatures higher than $\log\,T$=7.1; the
opposite trend is observed and essentially rules out this hypothesis.
\citet{Marino03} reached a similar conclusion for the supersaturated
star VXR45 based on an XMM-Newton observation.

Since the stars in our sample do not exhibit obvious supersaturation
characteristics, any clear significance of the results in the context
of other hypotheses attempting to explain the supersaturation
phenomenon is not obvious.  The rate of magnetic energy dissipation in
the coronae of our saturated and supersaturated stars, as gauged by
the X-ray to bolometric luminosity, is essentially the same; rather, it
is a fairly subtle difference in the properties of this
dissipation that gives rise to different coronal temperatures.

We also note that rapid rotation can give rise to a significant
reduction in the effective surface gravity at lower stellar latitudes;
in this context, the shift in plasma toward cooler temperatures in the
supersaturated regime goes contrary to expected scaling of increasing
coronal temperature with decreasing surface gravity for hydrostatic
models \citep[e.g.][]{Jordan91}.  However, direct X-ray spectroscopic
evidence based on {\it Chandra} HETG observations of AB~Dor suggests
that very active stellar coronae are indeed pole-dominated, as has
been hinted by the presence of large polar starspots
\citep{Drake07}, in which case the effects of centrifugal forces
would be lessened to some degree.

One further plausible culprit for influencing coronal plasma
temperature in otherwise similar stars is the chemical composition
through its influence on the radiative loss function.  As we discuss
below, there are some quite wide variations in composition within the
sample of stars condidered.  However, none of these appear to show a
significant correlation with coronal temperature past the
saturation-supersaturation boundary.

\subsection{Coronal Abundances}
\label{s:abundances}

In order to search for trends in coronal abundances with activity
parameters, we have also compared results derived here with other
results obtained for slightly ``slower'' rotators by ourselves and
other workers.
These results are summarised in Table~\ref{t:sample}.  Elements
considered are O, Mg, Si and Fe.  While other elements, such as S and
Ar, are also very interesting for abundance studies based on FIP or
other parameters, the He-like and H-like ions are not easily detected
in cooler coronae and it is difficult to accrue a meaningful sample
from the current literature.  We do not discuss Ne here as we found
its behaviour to follow closely that of O in the star sample
considered here; this is consistent with the study of
\citet{Drake05b} who found that a ratio Ne/O$\sim 0.4$ by number
represented well a sample of 21 mostly active stars.

All abundances are expressed
relative to the solar mixture recommended by \citet{Asplund05}, in the
usual spectroscopic bracket notation.  In the following discussion it
should, as usual, be borne in mind that the underlying stellar
photospheric abundances can differ substantially from solar values;
more important are differences relative to the solar {\em mixture} (ie
cases where, for elements A and B, [A/B]$ \neq 0$).  We note cases
where this is relevant in the apppropriate points in the discussion below.

\subsubsection{[Fe/H]}

\begin{figure}[t]
\plotone{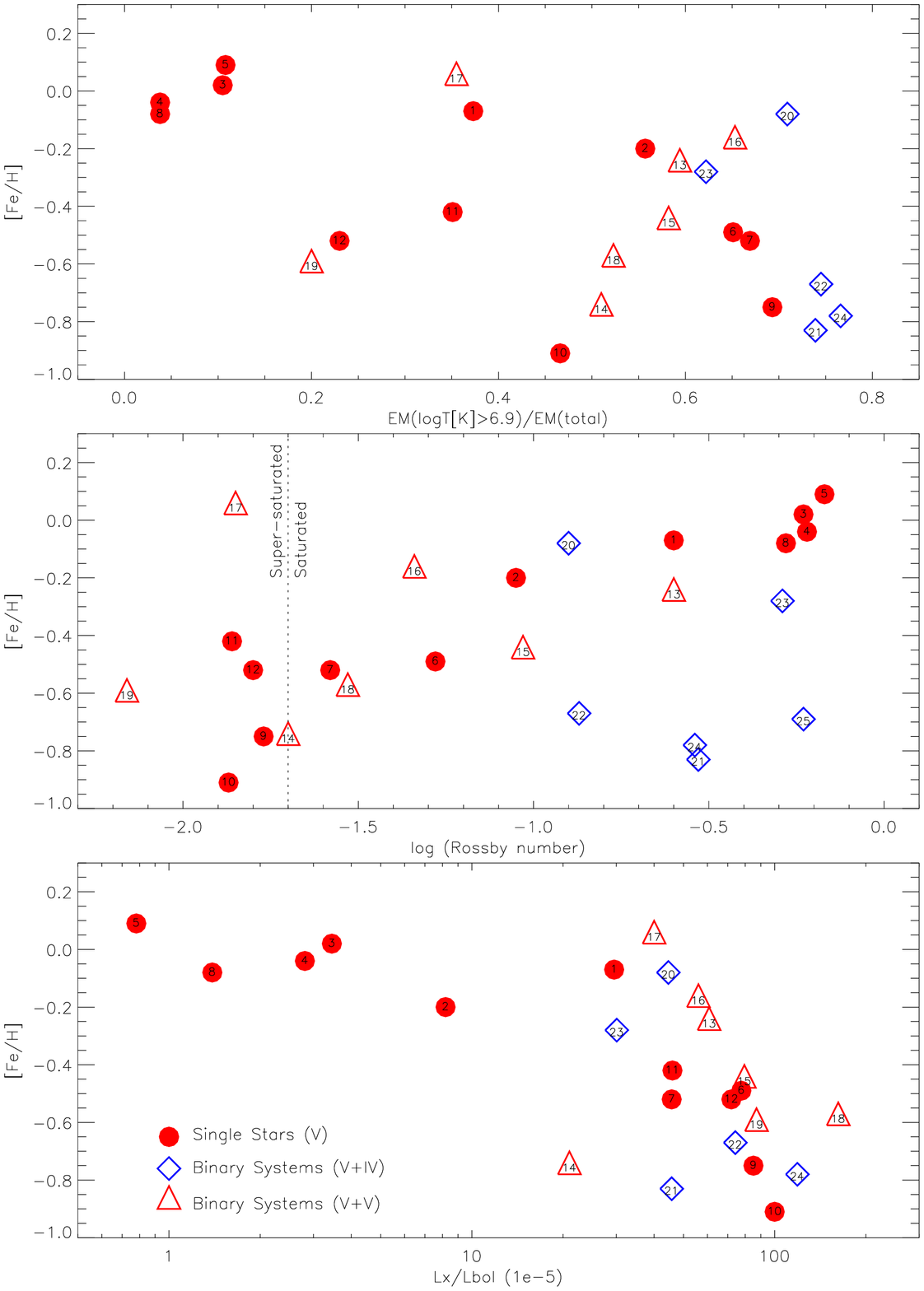}
\vspace{.6cm}
\caption{Top panel: [Fe/H] coronal abundance vs $\Phi_{6.9}$ for the sample of stars in Table~\ref{t:sample}. This sample includes single dwarfs (circles) stars and binary systems (diamonds and triangles). The coronal abundance values are expressed relative to the abundance mixture of \citet{Asplund05} (see discussion in \S4.4). Middle panel: Same as top panel but for the [Fe/H] coronal abundance vs the Rossby number. The demarcation between the saturated and supersaturated regions, based on definition of \citet{Randich98}, is indicated by a dashed vertical line.  Bottom panel: Same as top panel but for the [Fe/H] coronal abundance vs $L_x/L_{bol}$. The errors in the coronal abundances are $\sim$ 0.1 dex. Details on methods used for the
formal error estimates are given by \citet{Garcia-Alvarez05}}
\label{f:feh}
\end{figure}

The Table~\ref{t:sample} sample confirms nicely the trend of coronal
[Fe/H] with activity indices.  A relationship between Fe abundance and
activity was first suggested by \citet{Drake97} and \citet{Drake98} 
based on early {\it ASCA} and {\it EUVE} results, and has since been further 
fleshed-out based on {\it Chandra} and {\it XMM-Newton} observations of 
varied samples \citep[e.g.][]{Audard03,Guedel04}, as well as a study of
solar-like stars covering a range of activity
\citep{Telleschi05}.  In Figure~\ref{f:feh} we illustrate coronal
[Fe/H] for the sample as a function of $L_X/L_{bol}$, Rossby number,
and the DEM temperature-sensitive index $\Phi_{6.9}$.  The latter
relation shows an obvious, albeit scattered, correlation of
decreasing [Fe/H] with increasing hot emission measure fraction, as
expected based on the earlier work cited above.  Coronal temperature
might be regarded only as a proxy indicator of stellar activity---it is
observed to correlate well with $L_X/L_{bol}$ for example
\citep{Schmitt90}---and the trends of [Fe/H] with $L_X/L_{bol}$
and Rossby number are indeed more direct and interesting.

The relation with $L_X/L_{bol}$ shows a clear trend of declining iron
abundance with increasing activity level, with a break in behaviour at
$L_X/L_{bol}\sim 4\times 10^{-4}$ between shallow and steep declines
at low and higher activity levels, respectively.  Can any of this
trend be attributed to underlying stellar Fe abundances?  Among thin
disk stars, age metallicity effects would act in the opposite
direction to that observed: younger, more active stars should have
higher, not lower, metal abundances.  Moreover, the age spread for
stars considered here (a few Gyr or less) is also small for large
metallicity differences to be present.  It is possible, however, that
some of the binaries could belong to the thick disk population for
which lower metallicities would not be unusual.

In the crowded $L_X/L_{bol}> 4\times 10^{-4}$ region, stellar
metallicity might also be expected to play a role in the large
variation of coronal [Fe/H].  However, the strong trend here toward
lower coronal metallicity with higher $L_X/L_{bol}$ is shared among
single stars and binary stars of both evolved and unevolved type.  Of
the single dwarfs, noteable objects with large differences in coronal
[Fe/H] are AB~Dor and its companion Rst~137B, and the group of
intermediate activity stars $\kappa^1$~Cet, $\epsilon$~Eri,
$\chi^1$~Ori and $\pi^1$~UMa.  All these stars have measured
photospheric metallicities within $\sim 0.1$~dex of solar (see
Table~\ref{t:phot_abuns}).  In the middle of the range lie the young
Pleiades Moving Group G dwarfs 47~Cas~B and EK~Dra, whose photospheric
compositions are not expected to deviate from local cosmic values.
EK~Dra then appears to have an unmodified coronal Fe abundance, while
that of 47~Cas~B might be very slightly depleted.  The factor of 3
difference in coronal [Fe/H] of these stars {\em cannot be attributed
  to photospheric abundance differences} and must be the result of
chemical fractionation in the outer atmospheres of the more active
stars in which Fe is quite strongly depleted.

The relation with Rossby number also supports the above conclusion.  A
similar correlation among active dwarfs was reported earlier by
\citet{Singh99} based on low resolution ASCA spectra.  The dwarf
single and binary stars follow the clear trend of decreasing Fe
abundance with increasing activity (decreasing Rossby number)
elucidated as a function of $L_X/L_{bol}$, but here four of the
RS~CVn-type binaries are separated from the group (HR~1099, UX~Ari,
VY~Ari and II~Peg).  It is tempting to ascribe this to lower
photospheric Fe abundances: II~Peg has been studied in detail by
\citet{Berdyugina98} and \citet{Ottmann98} 
who found [M/H]$=-0.4$, and $=-0.16$, repsectively, while the subgiant
of the HR~1099 binary appears quite Fe-poor in the Li study
by \citet{Randich94} (though this latter study estimated
[Fe/H]=0 for the G dwarf secondary).  However, \citet{Ottmann98}
found essentially solar Fe for VY~Ari.  
The reason for the separation of these RS~CVn-type stars from the
group in the [Fe/H] vs. Rossby number plot then appears to be the
result of the different convection zone parameters of these somewhat
evolved stars: while it works well for dwarfs, Rossby number is not a
fundamental scaling agent for the coronal Fe content of stars of
different luminosity class.

Beyond the supersaturation limit, there appears to be a greater
scatter in [Fe/H] among dwarfs than at larger Rossby number.  The data
appear to be suggestive of a sharper drop in [Fe/H] as this limit is
reached.  However, since the M dwarfs AD Leo, EV Lac, and YY Gem all
lie above the earlier type active K-dwarfs Speedy Mic and Rst 137B in
this plot, the spread here might also be attributed to spectral type,
and consequently to differences in convection zone parameters.

\subsubsection{[O/H] and [O/Fe]}

Search for trends in O abundance with stellar activity indicators
carries the potential for systematic errors correlated with the index
being examined.  The available O feature in stellar X-ray spectra are
the He-like and H-like resonance whose emissivity functions peak at
$\log$ T=6.3 and 6.5, respectively. As noted above, dominant plasma 
temperatures increase with increasing activity, and lie at
$\log T\sim 7$ in active stars.  Errors in understanding DEMs at lower
temperatures where significant O~VII and O~VIII 
flux can originate then have the potential to skew 
abundance-activity relations.

Additionally, owing to the different contributions to metal enrichment
of the Galaxy through its history, comparison of O and Fe abundances
is further complicated by unknown or uncertain intrinsic metallicity,
and even whether stars belong to thin or thick disk populations.
\citet{Bensby04} show that local thick disk stars have higher [O/Fe]
by $\sim 0.2$~dex than those of the local thin disk at moderate metal
deficiencies ($-0.6 < $[Fe/H]$ < -0.1$); superimposed on both is the
usual trend of increasing [O/Fe] with declining [Fe/H].  This should
not be a problem for our young, single stars, but must be borne in
mind when considering the active binaries whose ages are much less
certain.

\begin{figure}[t]
\plotone{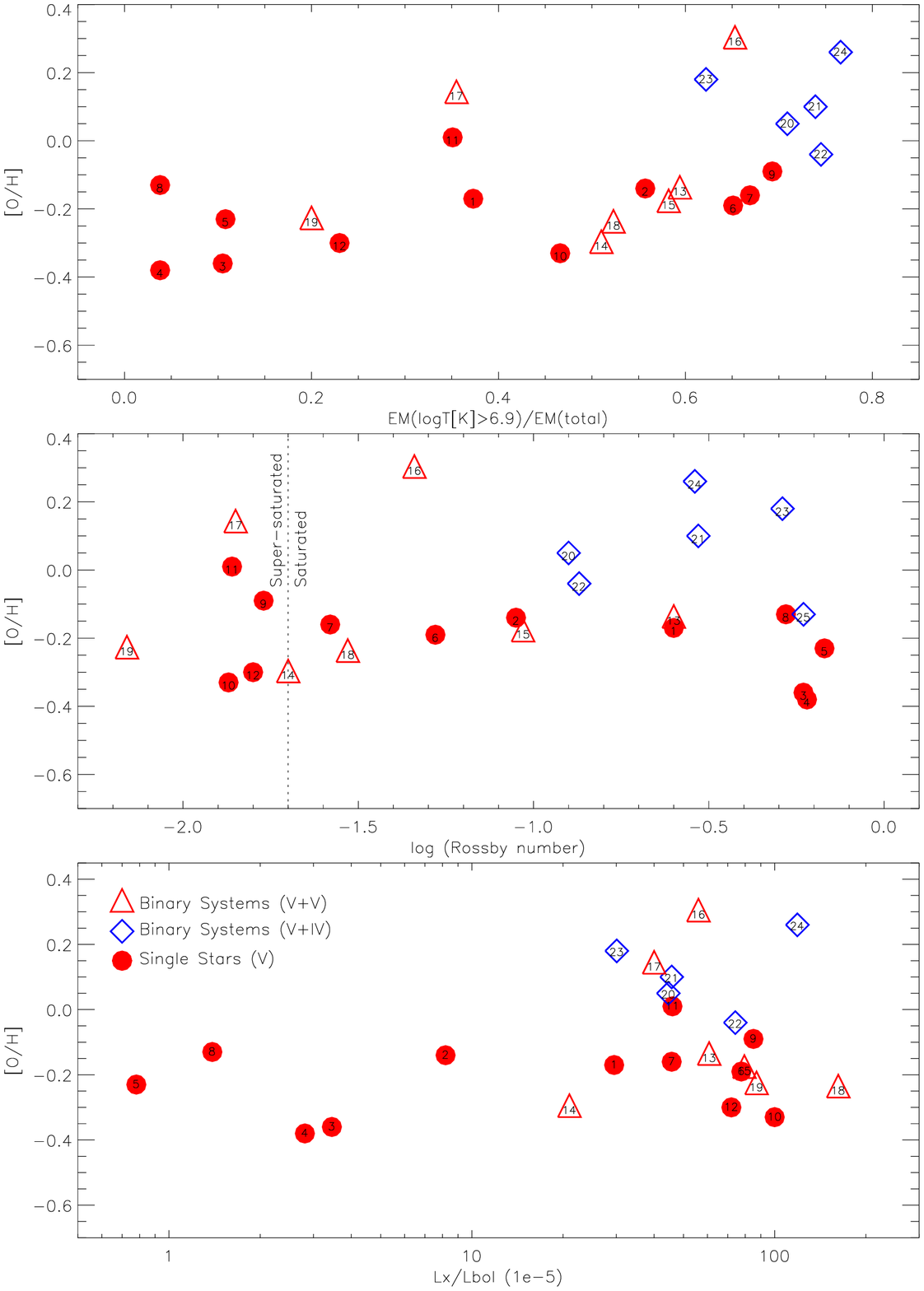}
\vspace{.6cm}
\caption{Same as Fig.~\ref{f:feh} but for [O/H].}
\label{f:ooh}
\end{figure}

We illustrate [O/H] as a function of the temperature index
$\Phi_{6.9}$, $L_X/L_{bol}$ and Rossby number in Figure~\ref{f:ooh}.
If the RS~CVn stars are excluded, there is a weak 
trend in the first of these of increasing [O/H] with
increasing $\Phi_{6.9}$.  A similar trend in Ne abundance is also
apparent in the literature survey of \citet{Guedel04}---as expected
based on the constancy of the Ne/O ratio noted earlier.  Is this
temperature trend a result of systematic error, underlying stellar
composition, or fractionation? The relation with $L_X/L_{bol}$ shows
the single stars to scatter about [O/H]$\sim -0.2\pm 0.1$, with the
RS~CVn-type binaries lying systematically above this; dwarf binaries
are instead scattered over a much larger range.  We
emphasise that we have referred the [O/H] values to the solar value
recommended by \citet{Asplund04}; were we to adopt that of
\citet{Grevesse98}, the above values would be $0.17$~dex {\em
lower}.  The photospheric O abundances for the G dwarfs $\chi^1$~Ori
and $\pi^1$~UMa are by differential analysis (ie essentially free from
modelling systematic errors) essentially solar 
yet have coronae apparently deficient in O by factors of more than 2;
it seems difficult to escape the conclusion that O must be depleted in
the coronae of these stars.  Similar conclusions were reached by
\citet{Wood06} in a detailed study of lower activity K dwarfs.

It is plausible that some of the range in [O/H] is due to underlying
stellar composition.  In the cases of II~Peg and V851~Cen,
photospheric evidence suggests mild metal paucity.  Were these to be
thick disk objects of near-solar metallicity the survey of
\citet{Bensby04} suggests an abundance [O/H]$\sim 0.2$ would not be
too unreasonable.

\begin{figure}[t]
\plotone{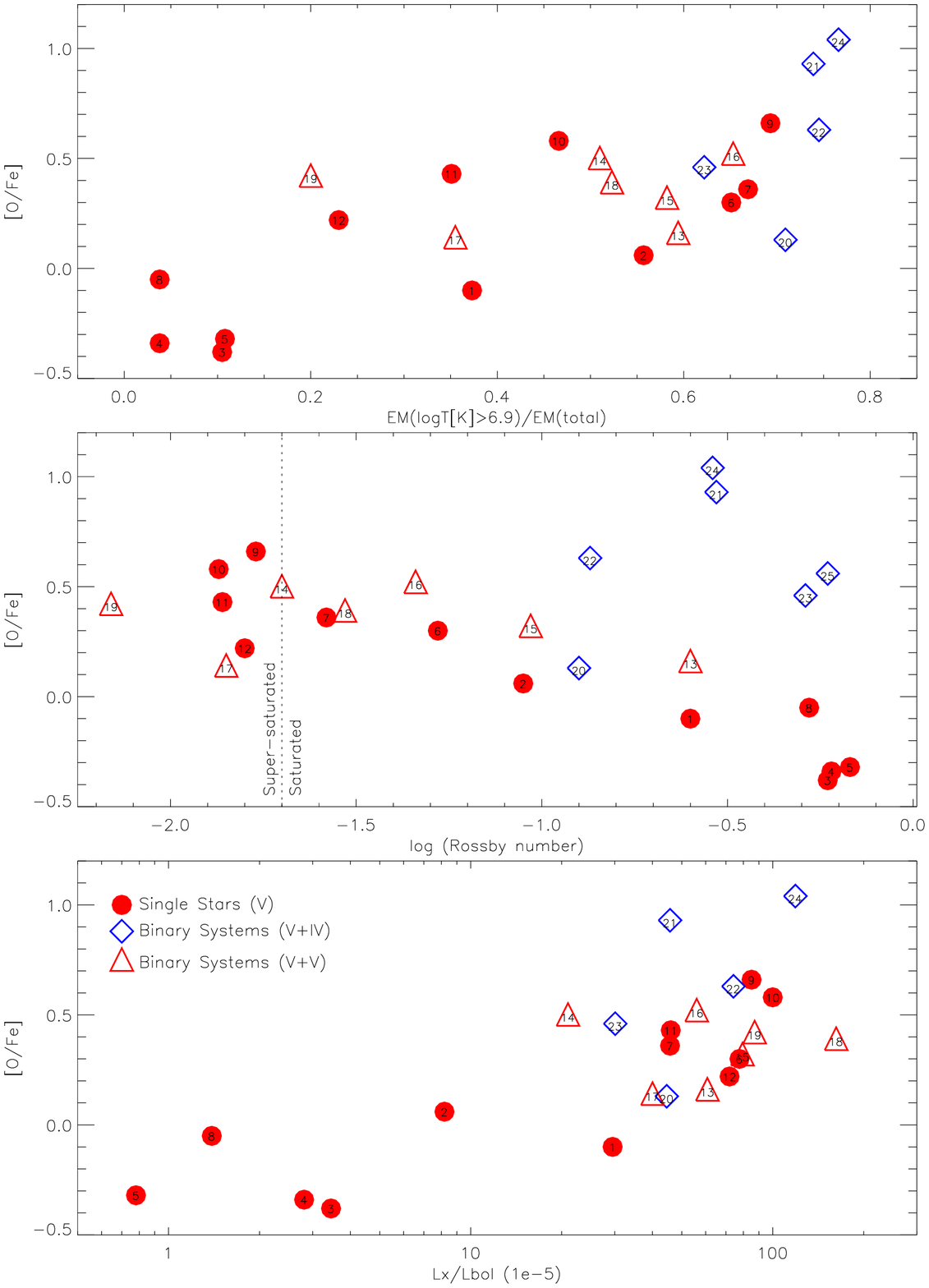}
\vspace{.6cm}
\caption{Same as Fig.~\ref{f:feh} but for [O/Fe].}
\label{f:oofe}
\end{figure}

The [O/Fe] ratio to some extent alleviates scatter due to global
metallicity differences between stars; these are illustrated in
Figure~\ref{f:oofe}.  The strong trends with all three activity
indices are readily apparent; similar trends in [O/Fe] with coronal
temperature were also noted for RS~CVn-types and a mixed sample by
\citet{Audard03} and \citet{Guedel04}, respectively.  Most
remarkable is the tight relation with Rossby number in which the
single and active binary dwarfs follow essentially the same increase
in [O/Fe] with decreasing Rossby number.  The trend is seen to break
at the supersaturation limit, beyond which there is no further
increase and some suggestion that [O/Fe] either saturates or perhaps 
declines slightly.

The RS~CVn-type binaries, with the exception of AR~Lac, show the same
conspicuous separation in [O/Fe] as they do in both [Fe/H] and [O/H],
only exacerbated by the ratio combination.  As a function of
$L_X/L_{bol}$, UX~Ari and II~Peg stand a factor of $\sim 2$ above the
scatter among the most active dwarf stars, and an order of magnitude
above the dwarfs of equivalent Rossby number.  We conclude that the
physical underpinning of these observations must depend strongly on
spectral type, and in particular on luminosity class. It is also possible 
that the different behavior of the RS~CVn-types is somehow a result of the influence 
of their binarity on chemical fraction processes.  That coronal plasma
temperature is not the key underlying factor is demonstrated by
the recent study of coronal abundances in giant stars by
\citet{Garcia-Alvarez06}: these stars have similar coronal
temperatures to the active dwarfs and RS~CVn-types, yet exhibit either
unmodified or low [O/Fe] ratios typical of a solar-like FIP effect.

If we exclude possible contributions to the observed range in coronal
O/Fe from intrinsic stellar scatter, coronal activity differences
among the sample appear to give rise to a variation in O/Fe by a
factor of 10.

While it is premature to speculate on fractionation mechanisms based
on the dataset considered here, it seems possible that the Alv\`en
wave pondermotive force model considered by \citet{Laming04} offers
some potential here.  \citet{Laming04} found that the fractionation
effect produced was quite strongly dependent on the wave propagation
and reflection properties of the chromosphere and lower corona, and on
the Alfv\'enic turbulence frequency spectrum---one might expect both
of these to be dependent on spectral type, rotation rate, and perhaps
also tidal distortion.

\subsection{[Mg/Fe] and [Si/Fe]}

Pursuit of these ratios is potentially important: any coronal
fractionation between the Mg-Si-Fe trio is interesting because these
elements have very similar FIPs.  Evidence that Fe appears
systematically low relative to Si and Mg in some stars was discussed
by \citet{Drake03a}, who suggested this as a possible diagnostic of
gravitational settling of the twice as heavy Fe relative to lighter Mg
and Si.  Acting against relative settling of heavier ions is mixing by
flows and turbulence, and kinetic effects such as thermal diffusion
that act in a temperature gradient to pull heavier ions toward hotter
temperatures.  Constraints on, or measurements of, gravitational
settling might then provide useful insights into coronal loop and
plasma flow properties.

As in the case of [O/Fe] noted above, ratios of the $\alpha$-elements
Mg and Si to that of Fe are also prone to scatter and systematic bias
as a result of Galactic chemical evolution.  Further systematic errors
are possible in the evaluation of the Si/Fe ratio owing to the
different temperatures over which the dominant Fe lines (weighted
largely by Fe~XVII at $\log T\sim 6.7$, though higher charge states
are often detected) and Si He-like and H-like lines ($\log T\sim 7.0-7.3$)
are formed.

Among our sample studied here, there are no obvious trends of Mg/Fe or
Si/Fe with activity parameters, except perhaps an indication of
declining Mg/Fe with coronal temperature index $\Phi_{6.9}$.  One
conspicuous outlier is II~Peg, which has both Mg/Fe and Si/Fe of
nearly three times the solar value; as noted earlier in the context
of its high O/Fe ratio, its mild photospheric metal deficiency
suggests that this might at least in part be explained by enhanced
photospheric $\alpha$-elements.  The ratios in the other stars are
also skewed toward values larger than unity, though by only 0.1~dex or
so.  This also could simply reflect photospheric values:
\citet{AllendePrieto04} note that the Sun appears deficient by
roughly 0.1 dex in O, Si, Ca, in addition to some other less abundant
elements compared to its immediate neighbors with similar iron
abundances.

Such a skew might also be due simply to systematic errors of
measurement rather than other mechanisms such as gravitational
settling.  However, as we noted earlier, we find Speedy Mic and
Rst~137B to have larger coronal Mg/Fe and Si/Fe ratios than AB~Dor
from a study involving the same lines and analysis techniques that
should minimise systematic errors.  These results suggest that some
fractionation is at work.

Disentangling all the possible contributions to the observed Mg/Fe and
Si/Fe ratios is well beyond our current study and requires more
detailed photospheric abundances.  More detailed investigation of
these ratios in a wider sample of stars would be well-motivated.

\section{Conclusions}
\label{s:concl}

AB~Dor, Speedy Mic and Rst~137B represent young ($\leq$ 100\,Myr)
rapidly rotating (P$_{orb}\leq$12\,hr) late-type stars lying at the
saturated-supersaturated corona boundary region.  Based on an analysis of high
resolution {\it Chandra} X-ray spectra of these stars and subsequent
comparison with results for other active stars culled from the
literature we draw the following conclusions.

\begin{enumerate}

\item The temperature structures of AB~Dor, Speedy~Mic and Rst~137B
all peak at $\log\,T$[K]$\sim$7.0-7.1, though the overall DEM shapes
are slightly different.  If the DEM trends observed here in only three
stars can be generalised, they hint that as supersaturation is reached
the DEM slope below the temperature of peak DEM becomes shallower,
while the DEM drop-off above this temperature becomes more
pronounced. 

\item In the context of the larger stellar sample, 
we observe that in dwarf single and binary stars 
coronal thermal structure shows an increase in
the emission of plasma at high temperatures ($\log T \ga 6.9$) 
as the Rossby number decreases and approaches the saturated-supersaturated
boundary.  However, once the
supersaturated region is reached this trend inverts; supersaturated
stars maintain a smaller fraction of coronal plasma at and above 10
million~K than stars of higher Rossby number.   This result is
consistent with the tentative generalised DEM behaviour outlined in
(1). 

\item All three of the stars studied in detail here 
show evidence for an inverse of the solar-like FIP 
effect, with smaller coronal abundances of the low FIP elements Mg, Si
and Fe, relative 
to the high FIP elements S, O and Ne.  This is consistent with
existing coronal abundance studies of other active
stars.

\item The stellar sample shows that 
coronal Fe abundance is inversely correlated with $L_X/L_{bol}$, and
for dwarfs is also 
well-correlated with Rossby number.  The Fe abundance is seen to decline
slowly with rising  $L_X/L_{bol}$, but 
declines sharply at $L_X/L_{bol}\ga 3\times 10^{-4}$.  

\item Coronal O abundances average at values of [O/H]$\sim -0.2$.
  Comparison of coronal and photospheric values for some of the sample
  suggests that active stellar coronae are in general slightly
  depleted in O relative to their photospheres. 

\item  The are no obvious trends of O abundance with activity
  indicators. Derived coronal O abundances are perhaps 
very weakly correlated with the coronal temperature
  index $\Phi_{6.9}$ with hotter coronae possibly 
exhibiting larger O abundances.  RS~CVn-type binaries exhibit
  systematically larger 
O abundances than dwarfs; this could be partially due to 
galactic evolutionary differences in [O/Fe] between dwarf and 
RS~CVn samples. 

\item The coronal O/Fe ratio for dwarfs shows a strong trend of increasing
  O/Fe with decreasing Rossby number, and appears to saturate at the
  supersaturation boundary with a value of 
  [O/Fe]$\sim 0.5$.  Similar correlations are seen with O/Fe
  increasing as a function of
  coronal temperature index, as revealed in earlier work, and with 
  increasing $L_X/L_{bol}$.  The range in O/Fe variations attributable
  to differences in coronal properties among the sample is about a
  factor of 10.

\end{enumerate}

\acknowledgments

DGA and WB were supported by {\it{Chandra}} grants GO1-2006X and
 GO1-2012X. LL was supported by NASA AISRP contract NAG5-9322; we
 thank this program for providing financial assistance for the
 development of the PINTofALE package. We also thank the CHIANTI
 project for making publicly available the results of their
 substantial effort in assembling atomic data useful for coronal
 plasma analysis.  JJD and VK were supported by NASA contract
 NAS8-39073 to the {\it{Chandra}}.
\clearpage


\newpage\section*{Tables}

\begin{table}[!h]
  \begin{center}
\caption{Summary of Stellar Parameters and HETG+ACIS-S Chandra
  Observations. 
\label{t:par}}
\begin{tabular}{lrccrccccccc}
\hline  \hline \\
Star & Sp.Tp. &dist.& B-V & $P_{rot}$ & T$_{eff}$  & $v$\,sin$i$ & log L$_{x}$/L$_{bol}$ &log N$_R$&ObsID&Date&Exp\\
&  & [pc] &  &[hr]  &[K]&   [km\,s$^{-1}$]  &&&&&[ks]\\
\hline \\
AB~Dor& K2V  &  14.9  & 0.80  & 12      &5250 & 93  &$\sim$-3&-1.58 &  16&1999-10-09&52.3\\
Speedy Mic  & K3V&44.4  & 0.94  & 9	&4750 & 132 &$<$-3 &-1.77&2536/3491&2002-04-26/27&35.0/35.1\\
Rst~137B     & M2V&14.9  & 1.60  & $<$9  &3250 & 50  &$\sim$-3&$<$-1.87&   16&1999-10-09&52.3 \\
\hline
\end{tabular}
\end{center}
\end{table}

\begin{table}
\caption{Identification and fluxes for spectral lines, observed on AB~Dor, Speedy Mic and Rst~137B, used in this analysis.\label{t:flx}}
\scriptsize{
\begin{center}
\begin{tabular}{lllccccl}
\hline  \hline \\
{$\lambda_{\rm obs}$} & 
{$\lambda_{\rm pred}$} &
{Ion} & 
{$log\,T_{\rm max}$} & 
{AB Dor}  & 
{Speedy Mic} &
{Rst~137B} &
{Transition (upper $\rightarrow$ lower)} \\
{[\AA]} & 
{[\AA]} &
{} & 
{[K]} & 
{[10$^{-14}$erg\,cm$^{-2}$\,s$^{-1}$]}  & 
{[10$^{-14}$erg\,cm$^{-2}$\,s$^{-1}$]} &
{[10$^{-14}$erg\,cm$^{-2}$\,s$^{-1}$]} &
 \\\\
\hline \\
\nodata &      4.733 & S     XVI  & 7.40 &$   2.2 \pm 0.8$&	\nodata     &  \nodata     &{\small $(2p) \; ^2P_{1/2}$ $\rightarrow$ $(1s) \; ^2S_{1/2}$}		 \\
     5.032 &      5.039 & S  XV   & 7.20 &$   8.8 \pm 3.0$&	\nodata     &  \nodata     &{\small $(1s2p) \; ^1P_{1}$ $\rightarrow$ $(1s^2) \; ^1S_{0}$}		 \\
     6.177 &      6.180 & Si XIV  & 7.20 &$  10.1 \pm 0.8$&$ 0.8 \pm0.2$&$0.3 \pm0.2$&{\small $(2p) \; ^2P_{3/2}$ $\rightarrow$ $(1s) \; ^2S_{1/2}$}		 \\
\nodata &      6.186 & Si    XIV  & 7.20 &$   5.0 \pm 0.4$&$ 0.4 \pm0.1$&$0.2 \pm0.1$&{\small $(2p) \; ^2P_{1/2}$ $\rightarrow$ $(1s) \; ^2S_{1/2}$}		 \\
     6.647 &      6.648 & Si XIII & 7.00 &$  13.5 \pm 1.1$&$ 1.1 \pm0.3$&$0.6 \pm0.3$&{\small $(1s2p) \; ^1P_{1}$ $\rightarrow$ $(1s^2) \; ^1S_{0}$}		 \\
     8.422 &      8.425 & Mg XII  & 7.00 &$   4.7 \pm 0.3$&$ 0.4 \pm0.1$&$0.2 \pm0.1$&{\small $(2p) \; ^2P_{1/2}$ $\rightarrow$ $(1s) \; ^2S_{1/2}$}		 \\
\nodata &      8.419 & Mg    XII  & 7.00 &$   9.4 \pm 0.6$&$ 0.8 \pm0.2$&$0.4 \pm0.2$&{\small $(2p) \; ^2P_{3/2}$ $\rightarrow$ $(1s) \; ^2S_{1/2}$}		 \\
     9.173 &      9.169 & Mg XI   & 6.80 &$   8.1 \pm 0.9$&$ 0.5 \pm0.2$&$0.5 \pm0.2$&{\small $(1s2p) \; ^1P_{1}$ $\rightarrow$ $(1s^2) \; ^1S_{0}$}		 \\
    10.998 &     11.003 & Na X    & 6.70 &$   8.2 \pm 1.2$&$ 0.5 \pm0.2$&$0.1 \pm0.1$&{\small $(1s2p) \; ^1P_{1}$ $\rightarrow$ $(1s^2) \; ^1S_{0}$}		 \\
    12.133 &     12.132 & Ne X    & 6.80 &$  72.5 \pm 1.9$&$ 9.2 \pm0.7$&$2.5 \pm0.5$&{\small $(2p) \; ^2P_{3/2}$ $\rightarrow$ $(1s) \; ^2S_{1/2}$}		 \\
\nodata &     12.137 & Ne    X    & 6.80 &$  36.2 \pm 1.0$&$ 4.6 \pm0.4$&$1.2 \pm0.2$&{\small $(2p) \; ^2P_{1/2}$ $\rightarrow$ $(1s) \; ^2S_{1/2}$}		 \\
    12.283 &     12.285 & Fe XXI  & 7.00 &$  22.7 \pm 2.1$&$ 1.6 \pm0.6$&$0.5 \pm0.4$&{\small $(2s^22p^3d) \; ^3D_{1}$ $\rightarrow$ $(2s^22p^2) \; ^3P_{0}$}	 \\
    13.433 &     13.447 & Ne IX   & 6.60 &$  49.5 \pm 2.7$&$ 4.0 \pm0.8$&$1.8 \pm0.7$&{\small $(1s2p) \; ^1P_{1}$ $\rightarrow$ $(1s^2) \; ^1S_{0}$}		 \\
    13.508 &     13.504 & Fe XIX  & 6.90 &$  19.5 \pm 2.3$&$ 0.3 \pm0.4$&$0.1 \pm0.1$&{\small $(2p^3(^2P)3d) \; ^1D_{2}$ $\rightarrow$ $(2s^22p^4) \; ^3P_{2}$}	 \\
    14.208 &     14.208 & Fe XVIII& 6.90 &$  10.5 \pm 0.8$&$ 0.5 \pm0.3$&$0.1 \pm0.2$&{\small $(2p^4(^1D)3d) \; ^2P_{3/2}$ $\rightarrow$ $(2s^22p^5) \; ^2P_{3/2}$}\\
\nodata &     14.203 & Fe    XVIII& 6.90 &$  19.8 \pm 1.5$&$ 1.0 \pm0.5$&$0.2 \pm0.3$&{\small $(2p^4(^1D)3d) \; ^2D_{5/2}$ $\rightarrow$ $(2s^22p^5) \; ^2P_{3/2}$}\\
    14.263 &     14.267 & Fe XVIII& 6.90 &$   4.4 \pm 1.0$&$ 0.3 \pm0.3$&$0.5 \pm0.3$&{\small $(2p^4(^1D)3d) \; ^2F_{5/2}$ $\rightarrow$ $(2s^22p^5) \; ^2P_{3/2}$}\\
    15.013 &     15.015 & Fe XVII & 6.75 &$  46.1 \pm 2.6$&$ 2.9 \pm0.8$&$1.5 \pm0.5$&{\small $(2p^53d) \; ^1P_{1}$ $\rightarrow$ $(2p^6) \; ^1S_{0}$}		 \\
    16.008 &     16.007 & O  VIII & 6.50 &$  12.4 \pm 0.8$&$ 0.8 \pm0.3$&$0.6 \pm0.2$&{\small $(3p) \; ^2P_{1/2}$ $\rightarrow$ $(1s) \; ^2S_{1/2}$}		 \\
\nodata &     16.006 & O     VIII & 6.50 &$  24.7 \pm 1.6$&$ 1.5 \pm0.6$&$1.2 \pm0.5$&{\small $(3p) \; ^2P_{3/2}$ $\rightarrow$ $(1s) \; ^2S_{1/2}$}		 \\
    18.973 &     18.973 & O  VIII & 6.50 &$  50.3 \pm 2.0$&$ 5.4 \pm0.9$&$2.7 \pm0.6$&{\small $(2p) \; ^2P_{1/2}$ $\rightarrow$ $(1s) \; ^2S_{1/2}$}		 \\
\nodata &     18.967 & O     VIII & 6.50 &$ 100.8 \pm 4.0$&$10.9 \pm1.8$&$5.4 \pm1.2$&{\small $(2p) \; ^2P_{3/2}$ $\rightarrow$ $(1s) \; ^2S_{1/2}$}		 \\
    21.607 &     21.602 & O  VII  & 6.30 &$  14.8 \pm 3.5$&$ 2.4 \pm1.8$&$1.5 \pm1.3$&{\small $(1s2p) \; ^1P_{1}$ $\rightarrow$ $(1s^2) \; ^1S_{0}$}		 \\
    24.777 &     24.779 & N  VII  & 6.30 &$  11.3 \pm 2.3$&	\nodata     &$0.1 \pm0.3$&{\small $(2p) \; ^2P_{3/2}$ $\rightarrow$ $(1s) \; ^2S_{1/2}$}		 \\
\nodata &     24.785 & N     VII  & 6.30 &$   5.6 \pm 1.2$&	\nodata     &$0.1 \pm0.2$&{\small $(2p) \; ^2P_{1/2}$ $\rightarrow$ $(1s) \; ^2S_{1/2}$}		 \\
\hline
\end{tabular}
\end{center}
}
\end{table}

\newpage

\begin{table}
\caption{Coronal abundances obtained from abundance-independent DEM-reconstructions.\label{t:abund1}}
\scriptsize{
\begin{center}
\begin{tabular}{lccccc}
\hline  \hline \\
Element\tablenotemark{a}& 
FIP\tablenotemark{b}&
AB~Dor&
Speedy Mic&
Rst~137B\\
\hline \\
$[$Na/H$]$&5.14 &$<$ -0.42\tablenotemark{c} &	\nodata    &   $<$-0.20\tablenotemark{c} \\
$[$Mg/H$]$&7.65 &-0.51 $\pm$   0.03&-0.70 $\pm$  0.07& -0.58 $\pm$   0.08\\
$[$Fe/H$]$&7.87 &-0.52 $\pm$   0.04&-0.75 $\pm$  0.07& -0.91 $\pm$   0.08\\
$[$Si/H$]$&8.15 &-0.46 $\pm$   0.03&-0.60 $\pm$  0.06& -0.63 $\pm$   0.07\\
$[$S/H$]$ &10.36&-0.32 $\pm$   0.11&    \nodata    &	 \nodata  \\
$[$O/H$]$ &13.62&-0.16 $\pm$   0.04&-0.09 $\pm$  0.08& -0.33 $\pm$   0.10\\
$[$Ne/H$]$&21.56&-0.25 $\pm$   0.03&-0.15 $\pm$  0.07& -0.49 $\pm$   0.08\\\\

\hline
\end{tabular}
\end{center}
\small{$^a$Logarithmic abundances relative to the abundance mixture of \citet{Asplund05} with Ne from \citet{Drake05b}.}\\
\small{$^b$First Ionization Potential in eV.}\\
\small{$^c$Upper limits due to lack of signal in line features.}\\
}
\end{table}

\begin{table}
\caption{AB~Dor, Speedy Mic and Rst~137B Abundance Ratios using
  Temperature-Insensitive Diagnostics.\label{t:abund2}} 
\begin{center}
\begin{tabular}{lccccccc}
\hline  \hline \\
Abundance&
AB~Dor&
AB~Dor\tablenotemark{a}&
Speedy Mic&
Speedy Mic\tablenotemark{a}&
Rst~137B&
Rst~137B\tablenotemark{a}\\
\hline \\
$[$N/O$]$&    +0.37$\pm$   0.09 & \nodata& \nodata            & \nodata& \nodata		 &\nodata\\
$[$O/Ne$]$&   -0.08$\pm$   0.11 &+0.09$\pm$0.07 & -0.01$\pm$   0.18  &+0.06$\pm$0.15 &+0.10$\pm$   0.19    &+0.16$\pm$0.18 \\
$[$Ne/Mg$]$&  +0.28$\pm$   0.04 &+0.26$\pm$0.06& +0.56$\pm$    0.16  &+0.55$\pm$0.14 &+0.05$\pm$   0.19    &+0.09$\pm$0.16\\
$[$Ne/Fe$]$&  +0.52$\pm$   0.03 &+0.27$\pm$0.07 & +0.80$\pm$   0.13  &+0.60$\pm$0.14 &+0.55$\pm$   0.17    &+0.42$\pm$0.16\\
$[$Mg/Si$]$&  -0.06$\pm$   0.04 &-0.05$\pm$0.06 & -0.08$\pm$   0.14  &-0.10$\pm$0.13 &+0.01$\pm$   0.26    &+0.05$\pm$0.15\\
$[$Mg/Fe$]$&  -0.19$\pm$   0.05 &-0.01$\pm$0.07 & -0.19$\pm$   0.28  &-0.05$\pm$0.14 &+0.21$\pm$   0.35    &+0.33$\pm$0.16\\
$[$Si/S$]$&   -0.14$\pm$   0.12 &-0.14$\pm$0.14 &	\nodata 	 &\nodata & \nodata	      &\nodata\\
\hline
\end{tabular}
\end{center}
\small{$^a$Abundance ratios using from the DEM-reconstructions.}\\
\end{table}

\begin{table}
\caption{Stellar Parameters and Coronal Abundances\tablenotemark{a}
  Single and Binary Systems
\label{t:sample}}
\scriptsize{
\begin{center}
\begin{tabular}{rlcrccccccccr}
\hline  \hline \\
{ID}&
{Single Star}&
{SpTy (V)}&
{P$_{rot}$}&
{log N$_R$}&
{log L$_x$/L$_{bol}$}&
{EM$_{low}$}&
{[Fe/H]}&
{[O/H]}&
{[Si/H]}&
{[Mg/H]}&
{[Ne/Fe]}&
{Ref}\\
\hline\\
 1&EK Dra          &G0V&2.75   &-0.60&-3.53&.627&-0.08&-0.10&-0.09&+0.11&+0.19& [15] \\
 2&47 Cas B	   &G1V&1.00   &-1.05&-4.09&.443&-0.21&-0.09&-0.14&-0.13&+0.43& [15] \\
 3&$\pi^{1}$ Uma   &G1V&4.70   &-0.23&-4.46&.895&+0.08&-0.22&-0.05&+0.21&-0.20& [15] \\
 4&$\chi^{1}$ Ori  &G1V&5.10   &-0.22&-4.55&.962&-0.01&-0.22&-0.10&+0.03&-0.04& [15] \\
 5&$\kappa^{1}$ Cet&G5V&9.20   &-0.17&-5.11&.892&+0.12&-0.05&+0.09&+0.34&+0.03& [15] \\
 6&PZ Tel 	   &K0V&0.94   &-1.28&-3.11&.349&-0.49&-0.19&-0.55&-0.46&+0.56& [1] \\
 7&AB Dor 	   &K2V&0.51   &-1.58&-3.34&.331&-0.52&-0.16&-0.46&-0.51&+0.67& [5] \\
 8&$\epsilon$ Eri  &K2V&11.03  &-0.28&-4.86&.962&-0.08&-0.13&-0.17&-0.10&+0.32& [13] \\
 9&Speedy Mic	   &K3V&0.38   &-1.77&-3.07&.307&-0.75&-0.09&-0.60&-0.70&+1.00&This work\\
10&Rst 137B	   &M2V&$<$0.37&-1.87&-3.00&.534&-0.91&-0.33&-0.63&-0.58&+0.82&This work\\
11&AD Leo 	   &M3V&2.70   &-1.86&-3.34&.649&-0.42&+0.01&-0.13&-0.34&+0.62& [10] \\
12&EV Lac 	   &M4V&4.38   &-1.80&-3.14&.773&-0.52&-0.30&-0.32&-0.61&+0.52& [12] \\

\hline  \hline\\
{ID}&
{Binary Syst.}&
{SpTy (V+V)}&
{P$_{rot}$}&
{log N$_R$}&
{log L$_x$/L$_{bol}$}&
{EM$_{low}$}&
{[Fe/H]}&
{[O/H]}&
{[Si/H]}&
{[Mg/H]}&
{[Ne/Fe]}&
{Ref}\\
\hline\\
13&TZ CrB   &F6V/G0V&1.14&-0.60&-3.22&.406&-0.28&-0.17&-0.05&+0.05&+0.34& [11] \\
14&44 Boo   &G0V/G5V&0.28&-1.70&-3.67&.490&-0.78&-0.33&-0.65&-0.67&+0.51& [3] \\
15&ER Vul   &G0V/G5V&0.69&-1.03&-3.10&.418&-0.48&-0.61&-0.49&-0.36&+0.75& [3] \\
16&SV Cam   &G0V/K6V&0.59&-1.34&-3.25&.347&-0.20&+0.27&-0.12&-0.04&+0.64& [14] \\
17&VW Cep   &G5V/K0V&0.25&-1.85&-3.40&.645&-0.67&-0.06&-0.43&-0.33&+0.46& [9] \\
18&V471 Tau &K2V/WD &0.52&-1.53&-2.79&.477&-0.61&-0.27&-0.46&-0.53&+0.62& [5] \\
19&YY Gem   &M1V/M1V&0.81&-2.16&-3.06&.800&-0.63&-0.26&-0.36&-0.65&+0.71& [6] \\ 
\hline  \hline\\
{ID}&
{Binary Syst.}&
{SpTy (IV+V)}&
{P$_{rot}$}&
{log N$_R$}&
{log L$_x$/L$_{bol}$}&
{EM$_{low}$}&
{[Fe/H]}&
{[O/H]}&
{[Si/H]}&
{[Mg/H]}&
{[Ne/Fe]}&
{Ref}\\
\hline\\
20&AR Lac   &G2IV/K0IV &1.98 &-0.90&-3.35  &.291   &-0.08&+0.05&-0.18&+0.16&+0.52& [8] \\
21&UX Ari   &K0IV/G5V  &7.44 &-0.53&-3.34  &.261   &-0.83&-0.30&-0.59&-0.54&+1.33& [3] \\
22&HR 1099  &K1IV/G5V  &2.84 &-0.87&-3.13  &.255   &-0.67&-0.04&-0.63&-0.72&+1.05& [4] \\
23&V851 Cen &K2IV-III/?&12.05&-0.29&-3.52  &.378   &-0.28&+0.18&-0.58&-0.07&+0.93& [13] \\
24&II Peg   &K2IV-V/?  &6.72 &-0.54&-2.93  &.234   &-0.78&+0.26&-0.30&-0.34&+1.36& [7] \\
25&VY Ari   &K3IV-V    &13.20&-0.23&\nodata&\nodata&-0.13&-0.69&-0.58&-0.47&+0.96& [2] \\
\hline \hline
\end{tabular}
\end{center}
\small{$^a$Abundances relative to solar photospheric values (Asplund
  et al. 2005).\\
References: [1] \citet{Argiroffi04}; [2] \citet{Audard04};  [3] \citet{Ball07};
  [4] \citet{Drake01}; [5] \citet{Garcia-Alvarez05}; [6]
  \citet{Guedel01b}; [7] \citet{Huenemoerder01};
  [8] \citet{Huenemoerder03}; [9] \citet{Huenemoerder06};  [10] \citet{Maggio04};
  [11] \citet{Osten03}; 
  [12] \citet{Robrade05}; [13] \citet{Sanz-Forcada04}; [14] \citet{Sanz-Forcada06};
  [15] \citet{Telleschi05}.} 
}
\end{table}

\begin{table}
\caption{Stellar Photospheric Abundances\tablenotemark{a}}
\label{t:phot_abuns}
\scriptsize{
\begin{center}
\begin{tabular}{rlccccr}
\hline  \hline \\
{ID}&
{Single Star}&
{[Fe/H]}&
{[O/H]}&
{[Si/H]}&
{[Mg/H]}&
{Ref}\\
\hline\\
1&EK Dra	   &+0.13&\nodata&\nodata&\nodata& [9] \\
2&47 Cas B	   &(-0.05&\nodata&\nodata&\nodata& [7])\\
3&$\pi^{1}$ Uma    &+0.10&+0.02&+0.12&+0.13& [1] \\
4&$\chi^{1}$ Ori   &+0.04&+0.15&+0.18&+0.24& [1] \\
5&$\kappa^{1}$ Cet &-0.36&-0.03&-0.22&-0.26& [1] \\
6&PZ Tel	   &(-0.58&\nodata&\nodata&\nodata& [7]) \\
7&AB Dor	   &+0.10&\nodata&+0.07&+0.05& [11] \\
8&$\epsilon$ Eri   &-0.17&-0.05&-0.08&-0.20& [1] \\
9&Speedy Mic	   &(+0.03&\nodata&\nodata&\nodata& [7]) \\
10&Rst 137B	   &+0.10&\nodata&+0.07&+0.05& [4] \\
11&AD Leo	   &-0.11&\nodata&\nodata&+0.00& [6] \\
12&EV Lac	   &\nodata&\nodata&\nodata&\nodata&\nodata\\\\
\hline \hline\\
{ID}&
{Binary Syst.}&
{[Fe/H]}&
{[O/H]}&
{[Si/H]}&
{[Mg/H]}&
{Ref}\\
&
{(V+V)}&
&
&
&
&
\\
\hline\\
13&TZ CrB      &\nodata&\nodata&\nodata&\nodata&\\
14&44 Boo      &(-0.41&\nodata&\nodata&\nodata& [7]) \\
15&ER Vul      &(-0.29&\nodata&\nodata&\nodata& [7]) \\
16&SV Cam      &\nodata&\nodata&\nodata&\nodata&\\
17&VW Cep      &\nodata&\nodata&\nodata&\nodata&\\
18&V471 Tau    &+0.18&\nodata&+0.20&\nodata& [10] \\
19&YY Gem      &\nodata&\nodata&\nodata&\nodata&\\\\
\hline \hline\\
{ID}&
{Binary Syst.}&
{[Fe/H]}&
{[O/H]}&
{[Si/H]}&
{[Mg/H]}&
{Ref}\\
&
{(IV+V)}&
&
&
&
&
\\
\hline\\
20&AR Lac      &+0.01&\nodata&+0.10&+0.01& [3] \\
21&UX Ari      &(-1.33&\nodata&\nodata&\nodata& [7]) \\
22&HR 1099     &(-1.25&\nodata&\nodata&\nodata&  [7]) \\
23&V851 Cen    &-0.18&\nodata&-0.01&+0.10& [5] \\
24&II Peg      &-0.16&\nodata&-0.13&-0.13& [8] \\
25&VY Ari      &-0.03&\nodata&-0.04&+0.04& [8] \\
\hline \hline
\end{tabular}
\end{center}
\small{$^a$Abundances relative to solar photospheric values (Asplund
  et al. 2005).\\
References: [1] \citet{AllendePrieto04}; [2] \citet{DrakeSmith93};
  [3] \citet{Gehren99}; [4] \citet{King00}; [5] \citet{Morel03};
  [6] \citet{Naftilan92}; [7] \citet{Nordstrom04} [based on photometric
  indices and liable to much larger uncertainties, especially for
  binaries];  [8] \citet{Ottmann98}; [9] \citet{Rocha-Pinto04};
  [10] \citet{Varenne99}; [11] \citet{Vilhu87}. 
}
}
\end{table}

\newpage

\newpage

\newpage

\end{document}